\documentclass[a4paper,fleqn,usenatbib]{mnras}

\usepackage{newtxtext,newtxmath}

\usepackage[T1]{fontenc}
\usepackage{ae,aecompl}

\usepackage{graphicx}	
\usepackage{amsmath}	
\usepackage{amssymb}	
\usepackage{hyperref}	
\usepackage{color}

\title[Star formation in the Magellanic Bridge]{Structured star formation in the Magellanic inter-Cloud region}

\author[Mackey et al.]{A. D. Mackey$^{1}$\thanks{E-mail: dougal.mackey@anu.edu.au}, S. E. Koposov$^{2,3}$, G. S. Da Costa$^{1}$, V. Belokurov$^{2}$, D. Erkal$^{2}$,
\newauthor F. Fraternali$^{4,5}$, N. M. McClure-Griffiths$^{1}$ and M. Fraser$^{6}$
\\
$^1$Research School of Astronomy and Astrophysics, Australian National University, Canberra, ACT 2611, Australia\\
$^2$Institute of Astronomy, University of Cambridge, Madingley Road, Cambridge, CB3 0HA, UK\\
$^3$McWilliams Center for Cosmology, Department of Physics, Carnegie Mellon University, 5000 Forbes Avenue, Pittsburgh, PA 15213, USA\\
$^4$Department of Physics and Astronomy, University of Bologna, viale Berti Pichat 6/2, I-40127 Bologna, Italy\\
$^5$Kapteyn Astronomical Institute, Postbus 800, NL-9700 AV Groningen, the Netherlands\\
$^6$School of Physics, O'Brien Centre for Science North, University College Dublin, Belfield, Dublin 4, Ireland
}

\date{\today.}

\pubyear{2017}

\begin{document}
\label{firstpage}
\pagerange{\pageref{firstpage}--\pageref{lastpage}}
\maketitle

\begin{abstract}
We use a new contiguous imaging survey conducted using the {\it Dark Energy Camera} to investigate the distribution and properties of young stellar populations in the Magellanic inter-Cloud region. These young stars are strongly spatially clustered, forming a narrow chain of low-mass associations that trace the densest H{\sc i} gas in the Magellanic Bridge and extend, in projection, from the SMC to the outer disk of the LMC. The associations in our survey footprint have ages $\la 30$\ Myr, masses in the range $\sim 100-1200\,{\rm M}_\odot$, and very diffuse structures with half-light radii of up to $\sim 100$\ pc. The two most populous are strongly elliptical, and aligned to $\approx 10\degr$ with the axis joining the centres of the LMC and SMC. These observations strongly suggest that the young inter-Cloud populations formed {\it in situ}, likely due to the compression of gas stripped during the most recent close LMC-SMC encounter. The associations lie at distances intermediate between the two Clouds, and we find no evidence for a substantial distance gradient across the imaged area. Finally, we identify a vast shell of young stars surrounding a central association, that is spatially coincident with a low column density bubble in the H{\sc i} distribution. The properties of this structure are consistent with a scenario where stellar winds and supernova explosions from massive stars in the central cluster swept up the ambient gas into a shell, triggering a new burst of star formation. This is a prime location for studying stellar feedback in a relatively isolated environment.
\end{abstract}




\section{Introduction}
The Large and Small Magellanic Clouds (LMC and SMC) are the two most massive satellite galaxies of the Milky Way, and the closest example of an interacting pair. They are situated deep inside the gravitational potential of the Galaxy at distances of $\approx 50$ and $60$\ kpc, respectively \citep*{degrijs:14,degrijs:15}.  The close proximity and locally-unique properties of the Clouds mean that they constitute exceptional laboratories for studying a huge variety of astrophysical problems, including critical general questions related to stellar evolution, star formation, the behaviour of the interstellar medium, and the evolution of galaxies \citep[see e.g.,][]{nidever:17}. 

One of the defining features of the Magellanic system is a vast envelope of neutral hydrogen gas stretching between the LMC and SMC \citep[the {\it Magellanic Bridge}; e.g.,][]{kerr:57,hindman:63} and extending into a filamentary stream spanning more than $200$ degrees across the sky \citep[the {\it Magellanic Stream}; e.g.,][]{mathewson:74,putman:03,bruns:05,nidever:10}. It was long thought that these structures were the result of repeated gravitational interactions between the Clouds and the Milky Way \citep[e.g.,][]{gardiner:96} and/or ram-pressure stripping by a hot gaseous component surrounding the Galaxy \citep[e.g.,][]{mastropietro:05}. However, recent measurements of the proper motions of the LMC and SMC are largely incompatible with these ideas -- the Clouds are moving too fast to be on stable short-period orbits about the Milky Way, and indeed are most likely on their {\it first} close passage \citep{kallivayalil:06a,kallivayalil:06b,kallivayalil:13,besla:07}. In this case the origin of the Magellanic Bridge and Stream cannot be attributed to the influence of the Milky Way; instead these features are generated by gravitational tides from the LMC acting on the SMC during repeated close encounters between the two \citep[e.g.,][]{besla:10,besla:12,diaz:11,diaz:12}. 

Such interactions can also explain many of the peculiar characteristics of the stellar component in the Clouds \citep[e.g.,][]{besla:16} such as the striking off-centre bar in the LMC \citep{devau:72,vdm:01}, a remote arc of stars in the extreme outskirts of the LMC disk \citep{mackey:16}, the irregular morphology of the SMC including a ``wing'' feature that extends towards the LMC \citep{shapley:40} and exhibits a substantial line-of-sight depth \citep{nidever:13}, the apparent existence of stripped SMC stars residing in the outskirts of the LMC \citep{olsen:11}, the presence of a substantial population of intermediate-age stars in the inter-Cloud region \citep[e.g.,][]{noel:13,noel:15,skowron:14,deason:17}, tidal tails extending from the SMC \citep{belokurov:17}, and a ``bridge'' of ancient metal-poor stars that stretches from the SMC almost to the LMC and which is strongly misaligned with the peak of the H{\sc i} \citep{belokurov:17}. 

It has been known for several decades that, in addition to the components described above, the Magellanic inter-Cloud region hosts a population of very young stars that stretches from the eastern ``wing'' of the SMC to the western periphery of the LMC \citep[e.g.,][]{irwin:85,irwin:90,demers:99}. Although sparsely distributed, particularly on the LMC side of the inter-Cloud region, this young population is observed to be significantly spatially clustered \citep[e.g.,][]{grondin:90,demers:91,battinelli:92,demers:98,bica:15}. Recent wide-field surveys have shown that the young stars closely trace the peak of the gaseous H{\sc i} Bridge between the Clouds, departing by a few degrees only at the easternmost end \citep[e.g.,][]{dinescu:12,skowron:14,noel:15,belokurov:17}. This tight correspondence suggests that these stars have likely formed {\it in situ} out of the H{\sc i} gas in the Magellanic Bridge, rather than having been stripped from the SMC.

\begin{figure}
\begin{center}
\includegraphics[width=0.5\textwidth]{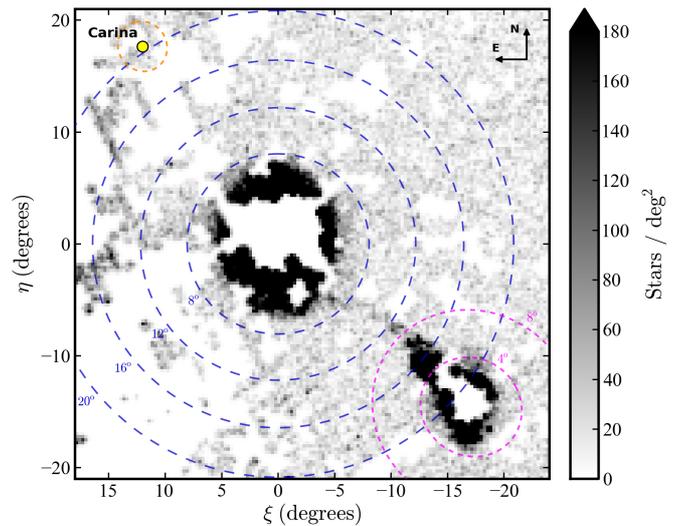}
\end{center}
\caption{Spatial density map of young (upper) main sequence stars in the Magellanic system, constructed using the ``GaGa'' {\it Gaia$+$GALEX} catalogue described by \citet{belokurov:17}. Stars were selected by requiring low foreground extinction $E(B-V) < 0.2$, and that $18 < G_0 < 20$ and $0 < (NUV-G)_0 < 1$ in colour-magnitude space. This map is a gnomonic projection with the centre of the LMC at the origin. The blue long-dashed circles indicate projected LMC-centric radii of $8\degr$, $12\degr$, $16\degr$, and $20\degr$ (or $\approx 7.0,\,10.5,\,14.0,$\ and $17.5$\ kpc), while the magenta short-dashed circles show projected distances of $4\degr$ and $8\degr$ from the SMC ($\approx 4.2$\ and $8.4$\ kpc). The distorted nature of the SMC is clearly visible in the form of its eastern wing; this feature extends into a narrow band of young stars that reaches to the south-western periphery of the LMC. It is also evident that, although diffuse, these populations appear rather spatially clustered. Gaps in the map largely result from incompleteness in the {\it GALEX} AIS catalogue \citep[see][]{belokurov:17}, although regions to the east of the map approach the Galactic plane and some data are consequently excised due to our cut on the colour excess. Sources near the eastern edge of the map are residual foreground contamination.
\label{f:gaga}}
\end{figure}

To highlight the properties of this young population, in Figure \ref{f:gaga} we map the spatial density of upper main sequence stars in the Magellanic system using the ``GaGa'' catalogue described by \citet{belokurov:17}. This consists of sources detected in both the {\it GALEX} ultraviolet space telescope's \citep{martin:05,morrissey:07} all-sky imaging survey \citep[AIS;][]{bianchi:14}, and the {\it Gaia} satellite's \citep{prusti:16} first data release \citep[DR1;][]{brown:16,lindegren:16}. We isolate young main sequence stars at the distance of the Magellanic Clouds by selecting only objects with $18 < G_0 < 20$ and $0 < (NUV-G)_0 < 1$ in colour-magnitude space\footnote{Here we are using measurements in {\it Gaia's} very broad optical $G$-band and {\it GALEX's} near-UV passband.}. To avoid issues due to regions of high extinction we further require that $E(B-V) < 0.2$; we de-reddened the photometry using extinction coefficients of $2.55$ for {\it Gaia} $G$ and $7.24$ for {\it GALEX} $NUV$. In Figure \ref{f:gaga} the distorted eastern wing of the SMC is clearly visible, together with a narrow band of young main sequence stars that extends from this structure across to the southern outskirts of the LMC. The clumpy nature of this young stellar bridge is also evident.

The brightest young stars in the inter-Cloud region are of significant importance as they allow various properties of the interstellar medium in the Magellanic Bridge to be measured that would otherwise be very difficult to access. Studies of absorption features along lines-of-sight to a handful of these hot stars (and a few background quasars) agree with high-resolution spectroscopy of the young stars themselves that the abundances of iron, as well as light elements such as C, N, O, Mg, and Si, are typically depleted in the Bridge by $\sim -1$ dex relative to the Milky Way \citep[e.g.,][]{hambly:94,rolleston:99,lehner:01,lehner:08,dufton:08,misawa:09}. Absorption measurements have further revealed that the interstellar medium in the Bridge region is highly complex, with molecules, neutral gas, and both weakly and highly ionised species all seen along various sightlines and at different velocities \citep[e.g.,][]{lehner:01,lehner:08,lehner:02}.

The observation that the Magellanic Bridge apparently has a lower abundance than either the LMC or SMC at the present day is puzzling; however it is in good agreement with similar measurements from numerous sightlines through the Magellanic Stream \citep[e.g.,][]{fox:10,fox:13}. The literature contains a variety of suggestions as to how this result might be reconciled with the general agreement of numerical models that the Magellanic Bridge consists of gas stripped from the SMC by a close encounter with the LMC -- for example, (i) the stripping may have occurred $\sim 1.5-2$ Gyr ago, when the abundance of the SMC was $\sim 0.1$ solar \citep{fox:13}; or (ii) in the case where the gas was stripped in the most recent interaction only $\sim 200$ Myr ago, there may have been significant dilution by a separate very low metallicity component \citep[e.g.,][]{rolleston:99,lehner:08}. In any case, it is clear that observations of the stars and gas in the Bridge region have the potential to provide important constraints on the interaction history of the LMC and SMC. They also offer insight into the nature of the star formation itself, which is notable because of its isolated location far out in the Milky Way halo, its low density, and its low metallicity. 

\begin{figure*}
\begin{center}
\includegraphics[height=85mm]{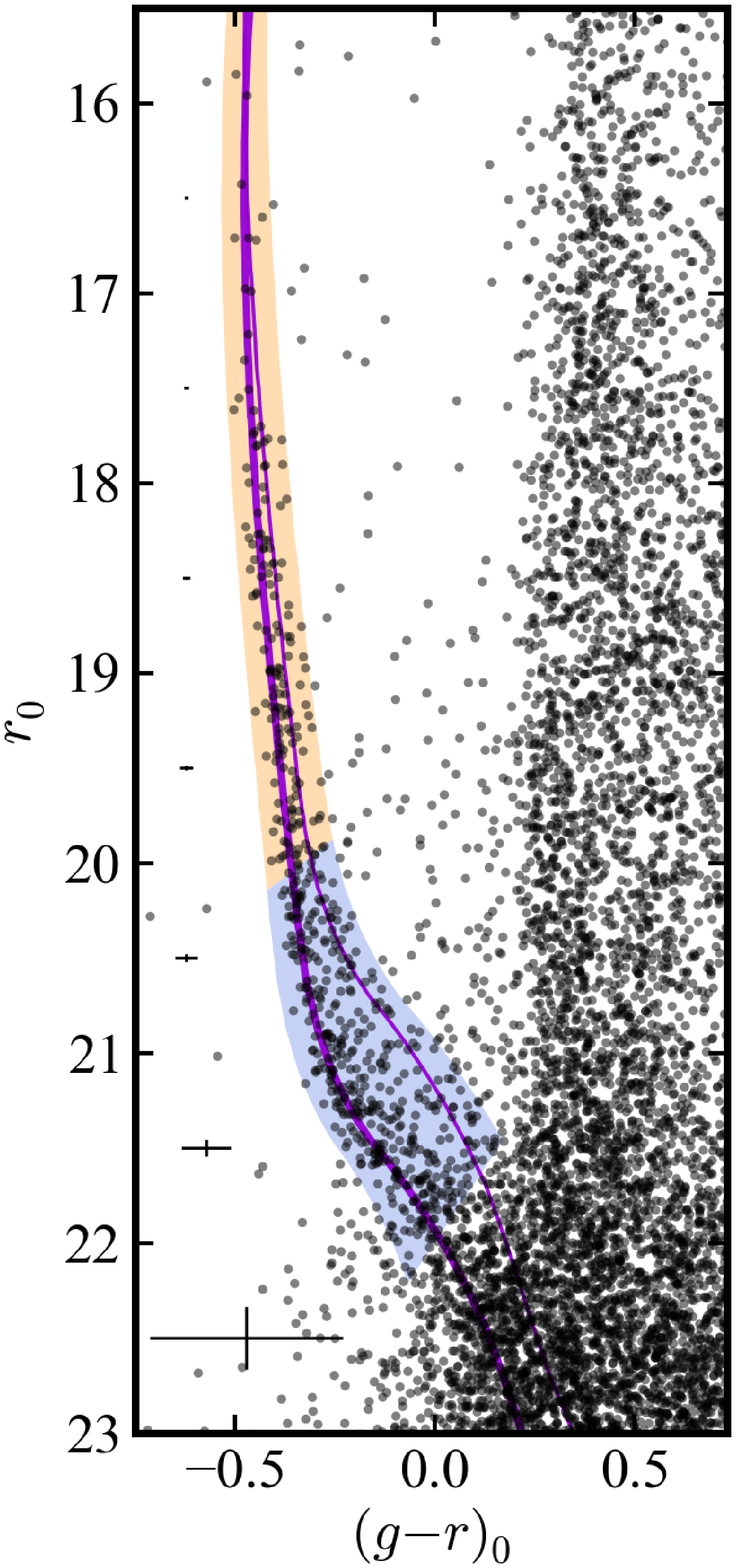}
\hspace{2mm}
\includegraphics[height=85mm]{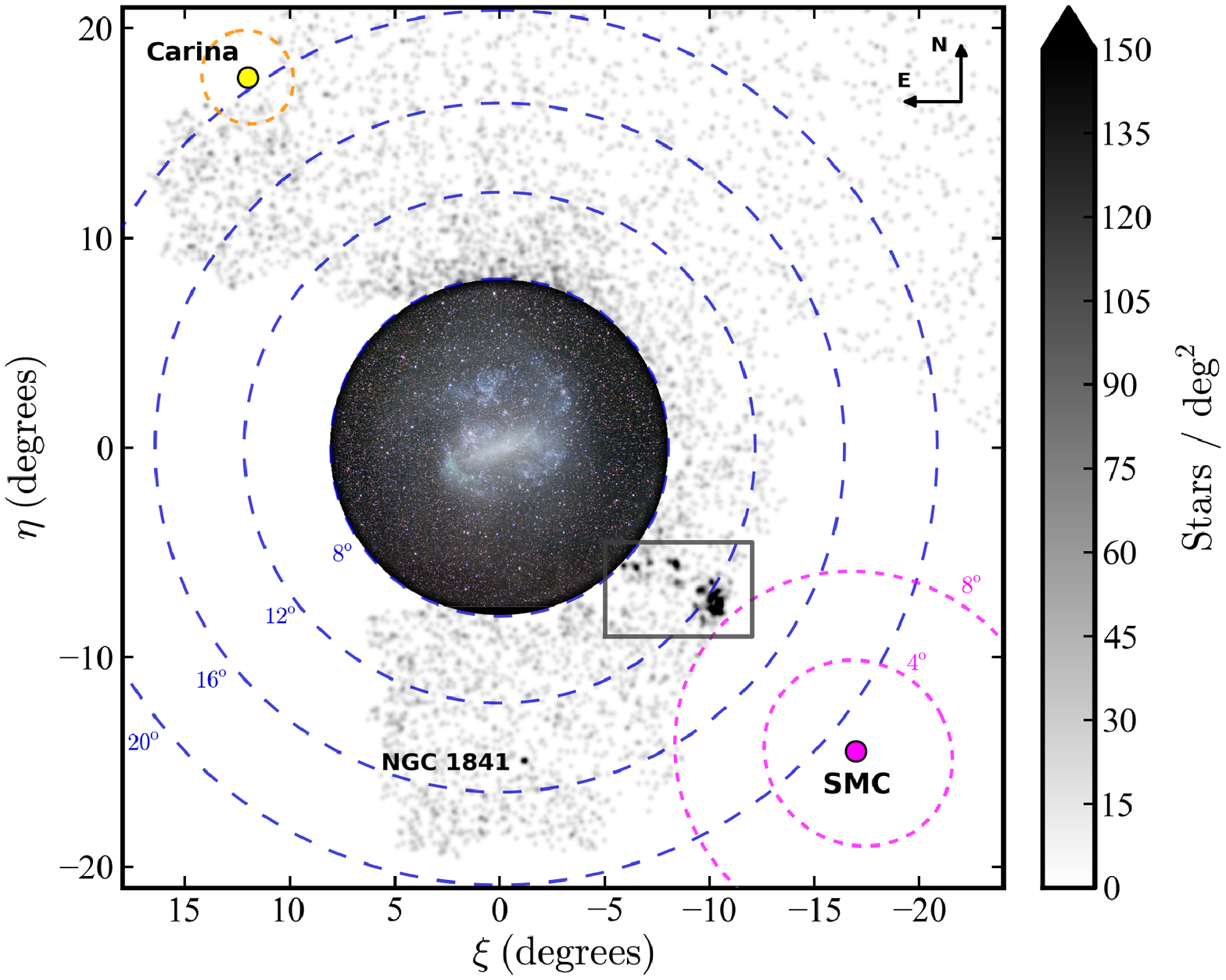}
\end{center}
\caption{{\bf Left:} Colour-magnitude diagram for stars lying at selected locations in the boxed region on the map to the right (see text). A substantial population of young blue stars is present, well separated from the Galactic foreground contamination that sits to the red. The thick solid line is a $30$\ Myr PARSEC isochrone \citep{bressan:12} with $[$M$/$H$] = -1.0$ and $\mu = 18.7$. The thin solid line is the equal-mass binary sequence for this fiducial track. The shaded region illustrates our main sequence selection box for the young population -- bounded by the isochrone to the blue and the binary sequence to the red, and broadened according to the photometric uncertainties at given $r_0$. The {\it upper} main sequence is delineated by the yellow portion of this box, extending down to $r_0 \sim 20$. Mean photometric uncertainties for stars with $(g-r)_0 < 0.0$ are plotted down the left-hand side of the panel. The mean $50\%$ completeness level sits at $r_0 \approx 23.0$. {\bf Right:} Spatial density map for upper main sequence stars across our complete survey footprint. The projection is the same as in Figure \ref{f:gaga}. As before, the blue long-dashed circles indicate LMC-centric radii of $8\degr$, $12\degr$, $16\degr$, and $20\degr$; inside $8\degr$ we display a wide-field optical image of the LMC (credit: Yuri Beletsky) to help place the size of the periphery into perspective. The locations of the SMC and the Carina dwarf are indicated, together with the LMC globular cluster NGC 1841; as before, the magenta short-dashed circles show distances of $4\degr$ and $8\degr$ from the SMC. The only significant concentration of young stellar populations within our survey footprint occurs in the inter-Cloud region.
\label{f:cmdmap}}
\end{figure*}

In this paper we present results from a new contiguous imaging survey that spans the eastern half of the Magellanic inter-Cloud region and reaches several magnitudes deeper than any extant mapping of this location with comparable spatial coverage. These data allow us to examine the spatial distribution of young stars in the Bridge with substantially higher contrast and resolution than previous studies (including Figure \ref{f:gaga}), as well as undertake the first detailed exploration of their relationship with the surrounding gas. 

\section{Observations and data reduction}
\label{s:data}
This work utilises the photometric catalogue presented by \citet{koposov:15} and \citet{mackey:16}, combined with new observations conducted by our group. All data were obtained using the {\it Dark Energy Camera} \citep[{\it DECam};][]{flaugher:15} mounted on the 4m Blanco Telescope at the Cerro Tololo Inter-American Observatory in Chile. {\it DECam} is a 520 megapixel imager consisting of 62 individual CCDs arranged into a hexagonal mosaic with a $\sim 3$\ deg$^2$ field of view.  The base catalogue is derived from publically-available images taken during the first year of the Dark Energy Survey \citep[DES;][]{abbott:05,abbott:06}, and covers the outskirts of the LMC to the north and north-west of the galaxy's centre. Our new data were obtained as part of program 2016A-0618 (PI: Mackey) over four half-nights on 2016 February 25-28, and span two separate regions adjoining the DES footprint. One covers $\approx 60$\ deg$^2$ and follows the extension of the stellar substructure discovered by \citet{mackey:16} to the north-east in the direction of the Carina dwarf, while the other covers $\approx 220$\ deg$^2$ of the LMC periphery to the west and south.  This latter area reaches roughly half-way to the SMC and thus spans approximately $50\%$ of the Magellanic inter-Cloud region. Our full survey footprint is displayed in the right-hand panel of Figure \ref{f:cmdmap} -- in the current work we focus on young stellar populations in the inter-Cloud region, while in an accompanying paper we present the discovery of several new low surface-brightness substructures in the outer LMC (Mackey et al. 2017, in prep.).

The entirety of our February 2016 observing run was clear and close to photometric. At each pointing we observed three $g$-band and three $r$-band frames. Exposure times were set to facilitate images matching the depth of those in the DES area to the north of the LMC; in practice, because of varying sky brightness (the moon, with $\sim 60\%$ illumination, set during each half-night) and seeing (many of our fields sit below $-75\degr$ declination) we employed integrations between $60-120$s per frame in $g$ and $40-80$s per frame in $r$. Across all images the median stellar FWHM is $1.20\arcsec$ in $g$ and $1.05\arcsec$ in $r$, each with rms scatter $0.15\arcsec$. All raw data were processed with the {\it DECam} community pipeline \citep{valdes:14}, and then passed through the photometry procedure described in detail by \citet{koposov:15}. Source detection and measurement was carried out with the {\it SExtractor} and {\it PSFEx} software \citep[e.g.,][]{bertin:96,bertin:11}, and a point-source catalogue was constructed by merging individual detection lists and removing galaxies using the {\sc spread\_model} and {\sc spreaderr\_model} parameters \citep[see][]{desai:12}. Photometry was calibrated to the SDSS scale using DR7 of the APASS survey and an ``\"{u}ber-calibration'' method \citep{pad:08}, and foreground reddening was corrected according to \citet*{schlegel:98} with the extinction coefficients from \citet{schlafly:11}. 

Following \citet{mackey:16} we estimated the detection completeness as a function of position by constructing the stellar luminosity function in $0.5\degr \times 0.5\degr$ bins and measuring the level of the turn-over; because we are only interested in the young Magellanic population in this paper, we restricted this process to use only stars with $(g-r)_0 < 0.0$. Across the inter-Cloud region the mean $50\%$ completeness level for such stars sits at $r_0 \approx 23.0$, although we note a general trend for the level to be shallower in the east (where it is typically $r_0 \approx 22.5$) and deepest in the west ($r_0 \approx 23.5$). This is partly a result of increased crowding in eastern fields, which lie closest to the LMC, and partly due to the variation in conditions (in particular the lunar illumination) under which the observations were conducted.

\section{Results}
\label{s:results}
Figure \ref{f:cmdmap} shows a map of our full survey area. Part of the inter-Cloud region is enclosed in a box; plotted alongside the map is a colour-magnitude diagram (CMD) for stars at selected locations in this area (specifically, this CMD constitutes the sum of those for the overdensities identified in Figure \ref{f:bridge}) . A substantial young population is clearly present.  Up to our saturation limit this is well traced by a PARSEC isochrone \citep{bressan:12} of age 30 Myr and $[$M$/$H$] = -1.0$, shifted to a distance modulus $\mu = 18.7$ (see Section \ref{ss:properties} and Figure \ref{f:youngcmds}). For bright stars, where the main sequence is nearly vertical, this model provides a good fit to the data; however, at fainter magnitudes (where the fiducial isochrone kinks to the red) the width of the main sequence is significantly broader than expected from the photometric uncertainties alone. The fiducial track for unresolved equal-mass binary stars, which sits\ $\approx0.75$\ mag above the single star isochrone \citep[e.g.,][]{hurley:98}, neatly encloses this broadening, suggesting that a substantial population of binaries is likely present.  This would be consistent with observations of young Magellanic Cloud clusters, where the binary fraction for mass ratios $q \ga 0.5$ is seen to be as high as $\sim 30-40\%$ \citep[e.g.,][]{elson:98,li:13}.  Together the single and binary star fiducial tracks define a CMD selection region, broadened by the observational uncertainties as a function of magnitude, for young Magellanic populations down to $r_0 \sim 22$. Note that the faint bound of this region sits well above the $50\%$ completeness level at all locations in our survey footprint. We use stars inside the upper part of the selection region ($r_0\,\la\,20$) to create the density map in Figure \ref{f:cmdmap}; this confirms that the only significant overdensity of young stellar populations within our survey footprint occurs in the boxed area.\footnote{The old LMC globular cluster NGC 1841 shows up in this map because it possesses an extended blue horizontal branch \citep[e.g.,][]{jeon:14} that overlaps with the young main sequence selection region on the CMD.}

\subsection{Properties of the young inter-Cloud populations}
\label{ss:properties}

\begin{figure}
\begin{center}
\includegraphics[width=0.5\textwidth]{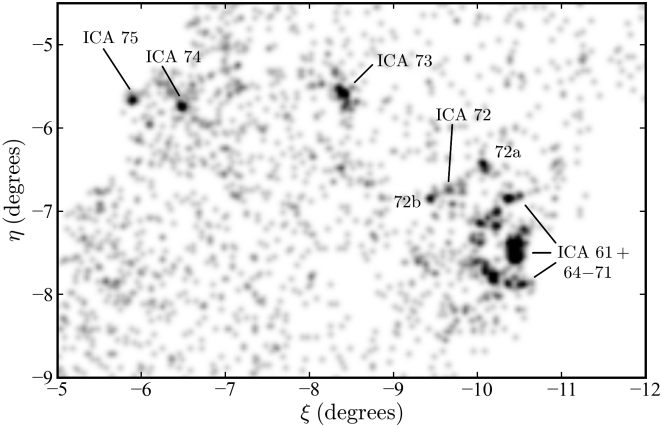}
\end{center}
\caption{Spatial density map in the eastern half of the inter-Cloud region for young stars isolated using the full CMD selection area from Figure \ref{f:cmdmap}.  Significant clustering of the young populations is evident; the associations catalogued by \citet{battinelli:92} are labelled. The striking core-shell structure described in the text sits near $(\xi,\,\eta) \approx (-10.5,\,-7.5)$.
\label{f:bridge}}
\end{figure}

Our new observations reach several magnitudes deeper than all previous surveys of the inter-Cloud region that have comparable spatial coverage and, as a consequence, we are able to trace the extent of the young populations with significantly greater contrast. Figure \ref{f:bridge} shows a spatial density map of the boxed inter-Cloud region from Figure \ref{f:cmdmap}, where young stars have been isolated using the full CMD selection region defined above. This reveals a narrow chain of young clusters and/or associations stretching from the western edge of our survey footprint (approximately $13\degr$ from the LMC and $9\degr$ from the SMC) to the outskirts of the LMC disk at a radius $\sim 8\degr$. Most of these features have previously been identified by \citet{battinelli:92}; in Figure \ref{f:bridge} we show the association names from this catalogue. However, the advantages conferred by the depth of our observations are evident.  For example, the structure labelled ICA 72 is in fact a much less significant density peak than two uncatalogued neighbouring features that we name ICA 72a and 72b. Also particularly striking is a ring or shell-like structure surrounding a compact core near $(\xi,\,\eta) \approx (-10.5,\,-7.5)$; $(\alpha,\,\delta) \approx (43.5,\,-73.5)$. Although this feature was briefly noted by \citet{irwin:90}, subsequent works have treated it as multiple independent young associations -- \citet{battinelli:92} identify eleven (of which two fall beyond the edge of our survey area).  In contrast, our deeper observations leave little doubt that the young populations at this location constitute a single vast, albeit undoubtably fragmented, structure\footnote{Again, the advantages of our deep imaging are clear -- of the seven associations identified by \citet{battinelli:92} that fall within our survey footprint and lie along the edge of the ring structure, only three match with obvious density peaks in our map. Moreoever, the four most significant density peaks along the edge of the ring are uncatalogued.}.

\begin{figure*}
\begin{center}
\includegraphics[height=100mm]{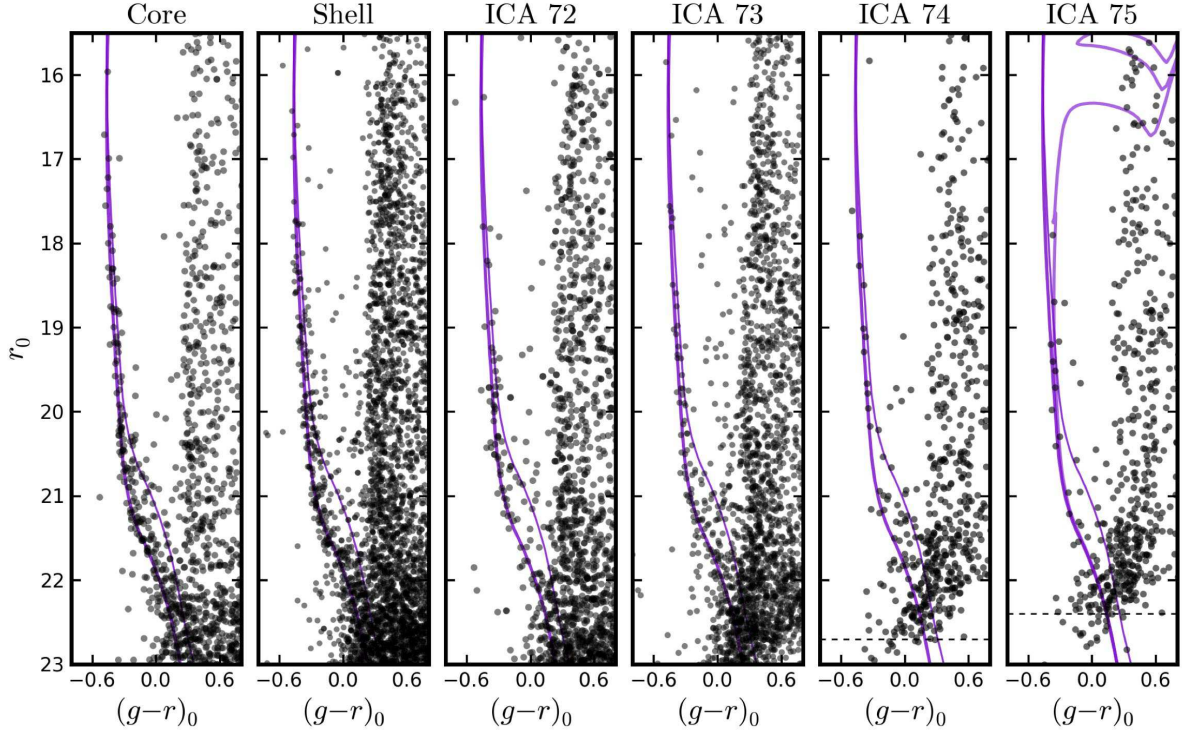}
\end{center}
\caption{Colour magnitude diagrams for six young associations in the eastern inter-Cloud region. We split the large structure at the western-most edge of our survey area into a ``core'' region (within $20\arcmin$ of its centre) and a ``shell'' region (stars in the range $20\arcmin-35\arcmin$ from its centre). The panel for ICA 72 includes all stars within $12\arcmin$ of the low-density structure identified by \citet{battinelli:92}, as well as stars within $9\arcmin$ of the previously-uncatalogued association we call ICA 72a and within $6\arcmin$ of the new association we call ICA 72b. The panels for ICA 73, 74 and 75 incorporate all stars within $20\arcmin$, $9\arcmin$, and $9\arcmin$ of the centres of these associations, respectively.  In each panel we plot a PARSEC isochrone with age $30$\ Myr and $[$M$/$H$] = -1.0$ (thick line) plus the equal-mass binary sequence (thin line), as shown in Figure \ref{f:cmdmap}. These are shifted to a distance modulus $\mu = 18.70$ for the core and shell, and $18.65$ for the other associations. The panel for ICA 75 also shows a $200$\ Myr isochrone with $[$M$/$H$] = -1.0$. In the panels for ICA 74 and ICA 75 we mark the level of the $50\%$ completeness limit with a horizontal dashed line; for the remaining four panels this level sits fainter than $r_0 = 23.0$.
\label{f:youngcmds}}
\end{figure*}
 
Figure \ref{f:youngcmds} shows CMDs for each of the main associations visible on our inter-Cloud map. The subplot labelled ``core'' refers to the compact cluster at the centre of the shell-like feature; the subplot labelled ``shell'' incorporates all stars in this feature (specifically, within $20\arcmin-35\arcmin$ from the centre of the core); and the subplot for ICA 72 includes all stars in both this structure and the two nearby higher density associations. The sampled areas are loosely scaled according to the spatial sizes of the associations (i.e., the half-light radii derived in Section \ref{ss:structures}) -- for each target we plot all stars within $2r_h$, and extend this boundary to larger radii in cases where the background contamination is not too severe. As in Figure \ref{f:cmdmap}, on each CMD we overlay a PARSEC isochrone of age 30 Myr and $[$M$/$H$] = -1.0$, shifted vertically to fit the observed stellar distribution. Our choice of metallicity is motivated by observations of gas and young stars in the inter-Cloud region, which, as described above, generally agree that the metallicity of the Magellanic Bridge is depleted by $\sim -1$ dex relative to the Milky Way \citep[e.g.,][]{hambly:94,rolleston:99,lehner:01,lehner:08,lehner:02,dufton:08,misawa:09}. 
 
\begin{table*}
\centering
\caption{Positions, ages, and distances for the main young associations in the eastern inter-Cloud region.}
\begin{minipage}{110mm}
\begin{tabular}{@{}lccccccc}
\hline \hline
Name & \multicolumn{2}{c}{Position (J2000.0)} & \multicolumn{2}{c}{Projected Coordinates} & Age & $\mu$ \\
 & $\alpha$ & $\delta$ & $\xi$ & $\eta$ & (Myr) & (mag)\vspace{1mm}\\
\hline
Core & $02^{\rm h}54^{\rm m}10\fs9$ & $-73\degr28\arcmin21\arcsec$ & $-10.45$ & $-7.50$ & $\approx 30$ & $18.70\pm0.05$\\
Shell & $\ldots$ & $\ldots$  & $\ldots$ & $\ldots$ & $\approx 15$ & $18.70\pm0.05$\\
ICA 72 & $03^{\rm h}08^{\rm m}53\fs0$ & $-73\degr18\arcmin48\arcsec$ & $-9.64$ & $-6.74$ & $\la 30$ & $18.65 \pm 0.05$\\
ICA 72a & $03^{\rm h}06^{\rm m}09\fs5$ & $-72\degr50\arcmin31\arcsec$ & $-10.08$ & $-6.45$ & $\la 30$ & $18.65 \pm 0.05$\\
ICA 72b & $03^{\rm h}10^{\rm m}25\fs2$ & $-73\degr30\arcmin11\arcsec$ & $-9.44$ & $-6.84$ & $\la 30$ & $18.65 \pm 0.05$\\
ICA 73 & $03^{\rm h}30^{\rm m}58\fs4$ & $-72\degr58\arcmin02\arcsec$ & $-8.39$ & $-5.59$ & $\la 30$ & $18.65 \pm 0.05$\\
ICA 74 & $03^{\rm h}53^{\rm m}44\fs6$ & $-73\degr55\arcmin49\arcsec$ & $-6.48$ & $-5.76$ & $\la 200$ & $18.65 \pm 0.05$\\
ICA 75 & $04^{\rm h}01^{\rm m}57\fs7$ & $-74\degr03\arcmin21\arcsec$ & $-5.89$ & $-5.66$ & $\la 200$ & $18.65 \pm 0.05$\\
\hline
\label{t:params}
\end{tabular}
\medskip
\vspace{-4mm}
\\
Notes: (i) Positions include the offsets ($x_0,y_0$) calculated as part of our structural measurements;\\
(ii) We list ICA 72 for completeness, even though it does not comprise a significant density peak;\\
(iii) We also list the two strong nearby density peaks that we label as ICA 72a and 72b (see text).
\end{minipage}
\end{table*}
 
\begin{table*}
\centering
\caption{Structures, luminosities, and masses for the main young associations in the eastern inter-Cloud region.}
\begin{minipage}{120mm}
\begin{tabular}{@{}lcccccccccc}
\hline \hline
Name & $r_h$ & $r_h$ & $e$ & $\theta$ & $N_{*}$ & $M_V$ & $M_{*}$ \\
 & (arcmin) & (pc) & & (deg) & & & (${\rm M_{\odot}}$)\vspace{1mm}\\
\hline
Core & $5.6^{+0.6}_{-0.5}$ & $89.5^{+9.6}_{-8.0}$ & $0.53\pm0.06$ & $37.6^{+5.4}_{-5.3}$ & $121\pm12$ & $-5.3^{+0.5}_{-1.1}$ & $900^{+110}_{-100}$ \\
Shell & $\approx 30$ & $\approx 480$ & $\ldots$ & $\ldots$ & $159\pm14$ & $-6.0^{+0.5}_{-0.9}$ & $1220^{+130}_{-120}$ \\
ICA 72a & $4.2^{+1.5}_{-1.3}$ & $65.6^{+23.4}_{-20.3}$ & $(0.00)$ & $\ldots$ & $24\pm7$ & $-3.1^{+0.7}_{-0.8}$ & $180\pm60$ \\
ICA 72b & $1.3^{+0.5}_{-0.3}$ & $20.3^{+7.8}_{-4.7}$ & $(0.00)$ & $\ldots$ & $11\pm3$ & $-2.1^{+0.9}_{-1.0}$ & $80\pm30$ \\
ICA 73 & $8.2^{+1.6}_{-1.3}$ & $128.1^{+25.0}_{-20.3}$ & $0.49\pm0.11$ & $51.7^{+9.9}_{-7.1}$ & $74\pm10$ & $-4.5^{+0.5}_{-1.5}$ & $540^{+90}_{-80}$ \\
ICA 74 & $0.8\pm0.2$ & $12.5\pm3.1$ & $(0.00)$ & $\ldots$ & $17\pm4$ & $-2.1^{+0.6}_{-1.1}$ & $130\pm40$ \\
ICA 75 & $1.8^{+0.6}_{-0.5}$ & $28.1^{+9.4}_{-7.8}$ & $(0.00)$ & $\ldots$ & $16\pm4$ & $-2.0^{+0.6}_{-1.1}$ & $120\pm40$ \\
\hline
\label{t:struct}
\end{tabular}
\medskip
\vspace{-4mm}
\\
Notes: (i) The quoted parameters are as follows: $r_h$ is the (elliptical) half-light radius; $e$ and $\theta$ are the ellipticity and position angle, respectively; $N_{*}$ is the number of member stars falling in our CMD selection box; and $M_V$and $M_{*}$ are the implied total luminosity and mass, respectively.\\
(ii) For the shell structure we list the measured radius in place of $r_h$.\\
(iii) The solutions derived for ICA 72a, 72b, 74, and 75 are constrained to have ellipticity $e=0$.
\end{minipage}
\end{table*}

Our photometry saturates above $r \sim 15.5$, so we are unable to resolve ages younger than $\approx 30$\ Myr using these CMDs; this represents a robust upper age limit for all the young associations plotted in Figure \ref{f:youngcmds} except for ICA 74 and ICA 75. These two objects are poorly populated and lack upper main sequence stars (at least up to our saturation limit) -- this could simply be due to stochastic sampling of the stellar mass function, or it might indicate that these two associations are older than the others. Formally our upper age limit for these two clusters is $\approx 200$\ Myr; an isochrone of this age is plotted on the CMD for ICA 75. 

Because our photometry reaches well below the bend in the main sequence between $r_0 \sim 21-22$, it is possible to obtain precise relative distances for the main aggregates plotted in Figure \ref{f:youngcmds}.  This has not previously been possible except for targeted deep observations of a handful of associations across the inter-Cloud region \citep[e.g.,][]{demers:98}. We find that the core-shell structure sits at a distance modulus of $\mu = 18.70 \pm 0.05$, while the ICA 72 complex, together with ICA 73, 74, and 75, are all slightly closer with $\mu = 18.65 \pm 0.05$ (here, all the quoted uncertainties are random). These measurements preclude a distance gradient greater than $\Delta\mu \approx 0.15$ mag across the eastern half of the inter-Cloud region. At low significance they are consistent with a mild gradient such that the objects sitting closer to the LMC in projection have shorter line-of-sight distances by $\approx 0.05$ mag than those sitting closer to the SMC in projection \citep[][obtained a similar result]{demers:98}; however the data are also formally consistent with a scenario where all the associations sit at the same distance.

Note that these conclusions assume that all the clusters plotted in Figure \ref{f:youngcmds} have approximately the same metallicity. We find that isochrones of age $30$\ Myr and $[$M$/$H$] = -0.5$ and $-1.5$ fit the data as well as that for our assumed $[$M$/$H$] = -1.0$, but require different vertical offsets. Changes of $\pm 0.5$ dex in metal abundance lead to changes in the measured distance modulus of roughly $\pm 0.2$ mag. However, the fact that we observe strong consistency between the distance estimates across our sample suggests that any cluster-to-cluster metallicity variations are probably much smaller than this. It is worth emphasizing that all the associations we have measured here, even those at easternmost edge of the inter-Cloud region, sit on the far side of the LMC \citep[for which we assume $\mu = 18.49$;][]{degrijs:14}.  For our adopted $[$M$/$H$] = -1.0$ the associations have an average line-of-sight distance $\approx 4$\ kpc greater than that to the LMC centre; however, their distance moduli are larger than $18.49$ for all metallicities in the range $-1.0 \pm 0.5$ (with the metal-poor end leading to distances approximately matching that of the LMC). This is in contrast to the results of \citet{bica:15} who measured young associations in the SMC half of the inter-Cloud region to lie {\it closer} to us than the LMC.

Table \ref{t:params} lists coordinates for each of the main associations discussed above, and summarizes our age and distance estimates.  Note that the entries for the core-shell structure reflect refined age estimates that we derive in Section \ref{s:discussion}.

\subsection{Structures and luminosities of the main associations}
\label{ss:structures}
It is informative to derive structural parameters, and luminosity and mass estimates, for each of main inter-Cloud associations in our survey. To achieve this we employ the methodology developed by \citet{martin:08,martin:16}. For a given target we select all $N$ stars in some local region $\mathcal{A}$ that fall inside the CMD selection box defined in Figure \ref{f:cmdmap}, and determine the density model:
\begin{equation}
\Sigma(r) = \frac{1.68^2}{2\pi r_h^2 (1-e)} N_* \exp(-1.68r/r_h) + \Sigma_b
\end{equation}
that maximises the likelihood of the data. This consists of a simple exponential radial decline in the surface density\footnote{While there are a wide variety of models that one might consider fitting to a cluster-like system, the objects studied here are sufficiently poorly populated that none provides any particular advantage over the others. The exponential models we adopt provide a convenient description of the data and utilise one fewer parameter than most other families; moreover, it is commonly observed that young star clusters in the Magellanic Clouds tend to exhibit radial surface density profiles that are not obviously truncated \citep*[e.g.,][]{eff:87,mackey:03a,mackey:03b}.}, plus a background level $\Sigma_b$ that is assumed constant over $\mathcal{A}$. This background is computed in terms of $N$ and the integral of the exponential profile over $\mathcal{A}$ \citep[see equation 6 in][]{martin:16}; we note that $\mathcal{A}$ need not be continuous, which is relevant because several of the target associations fall close to the edge of our survey footprint. The quantity $r$ is the elliptical radius, defined in terms of the projected sky coordinates $(x,y)$, an offset ($x_0,y_0)$ from the initial guess for the association centre \citep[for which we use the catalogue of][]{battinelli:92}, and the ellipticity $e = 1-b/a$ and position angle $\theta$ east of north \citep[see equation 5 in][]{martin:16}. The assumed exponential radial profile has a half-light radius $r_h$, while $N_*$ is the number of stars from the input list of $N$ that belong to the system.  
 
\begin{figure}
\begin{center}
\includegraphics[width=0.45\textwidth]{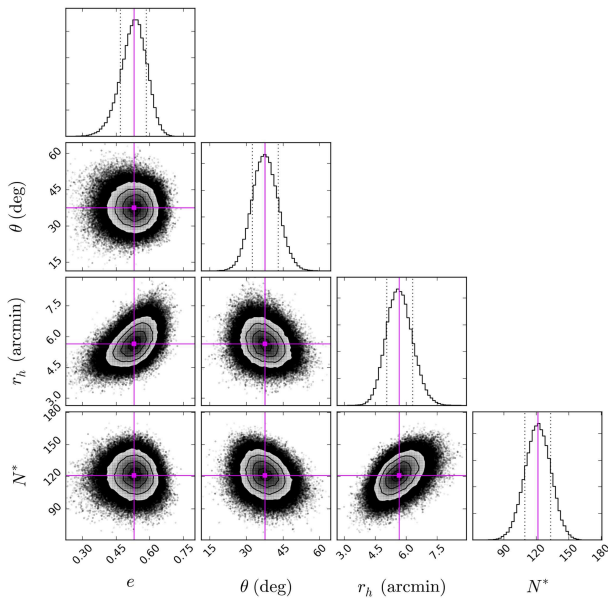}
\end{center}
\caption{Structural solution for the ``core'' association. Marginalized one- and two-dimensional posterior PDFs are shown for the four physically important parameters in our exponential radial density model. These are the ellipticity $e$, the position angle $\theta$ east of north, the half-light radius $r_h$, and the number of stars in the region $\mathcal{A}$ lying in the CMD selection box that belong to the system in this solution. The contours show the $1\sigma$, $2\sigma$, and $3\sigma$ confidence levels. Created with {\it corner.py} \citep{fm:16}.
\label{f:coremcmc}}
\end{figure}

We use the Markov chain Monte Carlo (MCMC) package {\it emcee} \citep{fm:13} to sample the posterior probability distribution functions (PDFs) and infer the most likely set of model parameters $\{x_0,y_0,e,\theta,r_h,N_*\}$ given our observations. We assume flat priors for all parameters and impose the following restrictions to ensure that the solutions remain meaningful: $0.0 \leq e < 1.0$, $0.0 \leq \theta < 180.0$, $r_h > 0.0$, and $0 < N_* \le N$. As an example outcome, we consider the solution for the ``core'' association. Figure \ref{f:coremcmc} shows the one- and two-dimensional marginalized PDFs for the four physically important parameters in our model, $\{e,\theta,r_h,N_*\}$. For this amply-populated system and for ICA 73, the solutions are well defined; however we found that the smaller associations ICA 72, 72a, 72b, 74 and 75 all possess too few stars for the algorithm to properly converge. In order to obtain some measure of the sizes of these systems we attempted to simplify the problem by fixing the ellipticity to be zero (and thus removing two parameters) -- this facilitated convergence in all cases except for ICA 72, which appears not to constitute a sufficiently significant density enhancement relative to the background (see below).
 
\begin{figure*}
\begin{center}
\includegraphics[width=80mm]{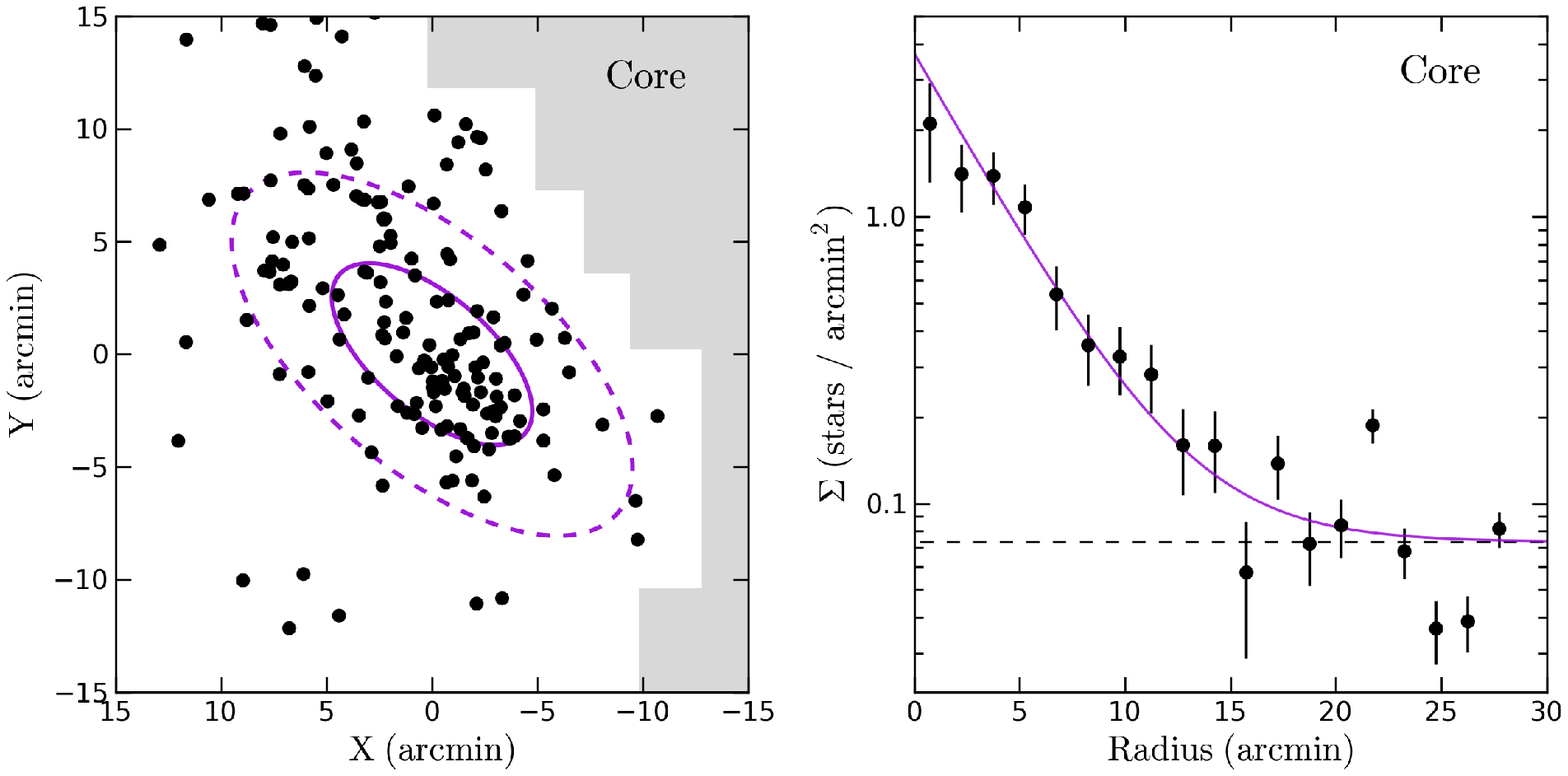}
\hspace{5mm}
\includegraphics[width=80mm]{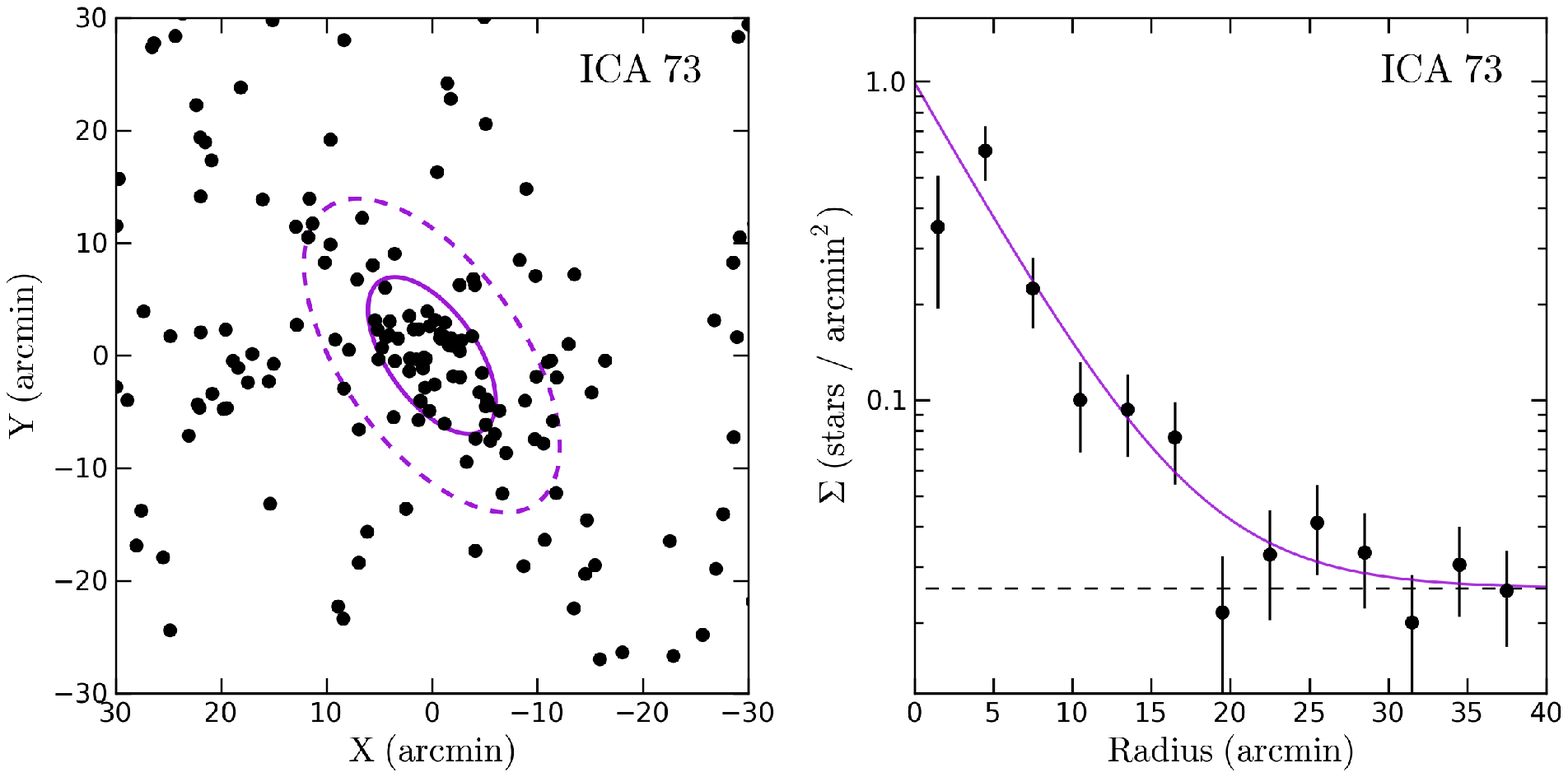} \\
\vspace{1mm}
\includegraphics[width=80mm]{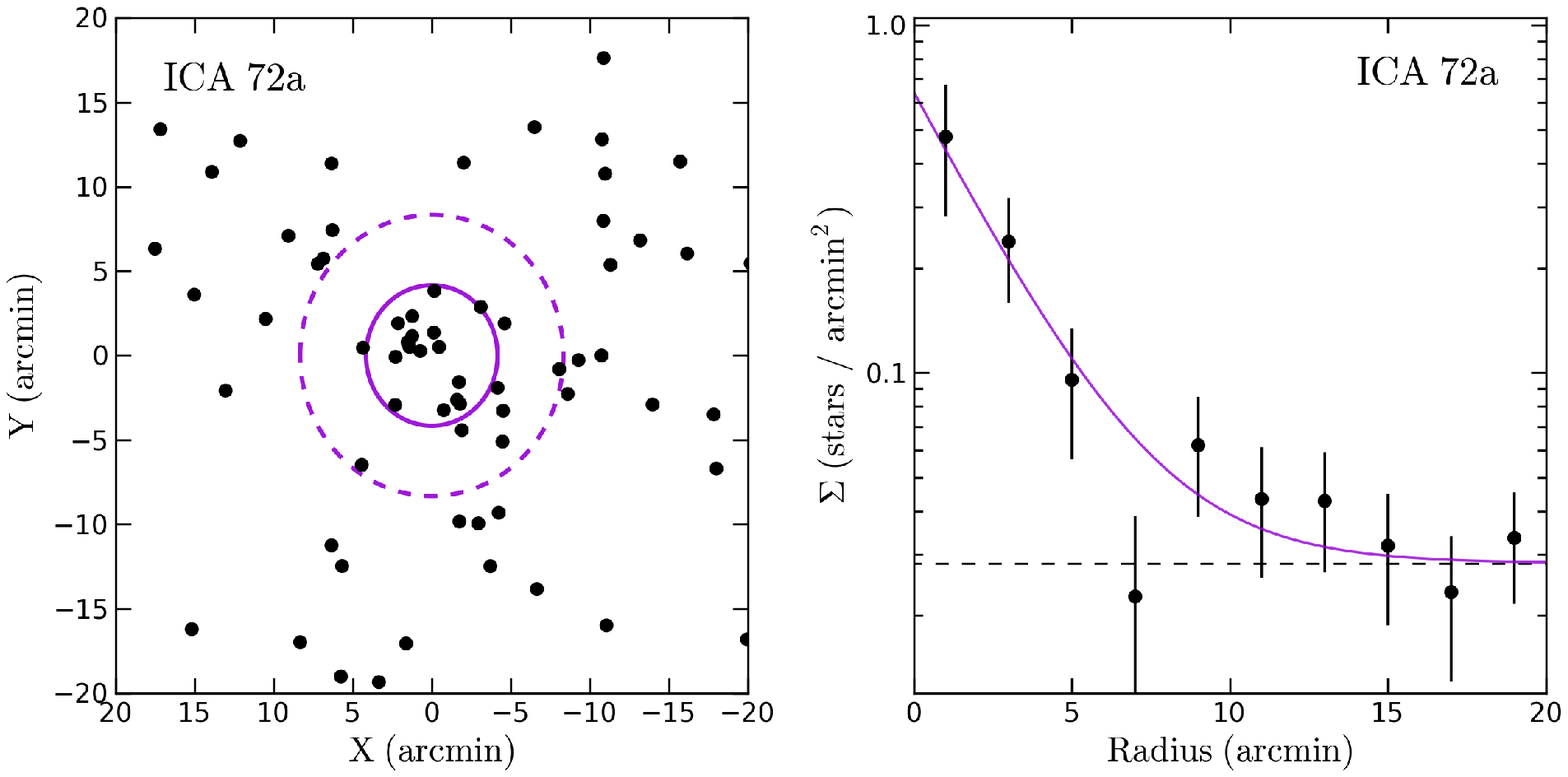} 
\hspace{5mm}
\includegraphics[width=80mm]{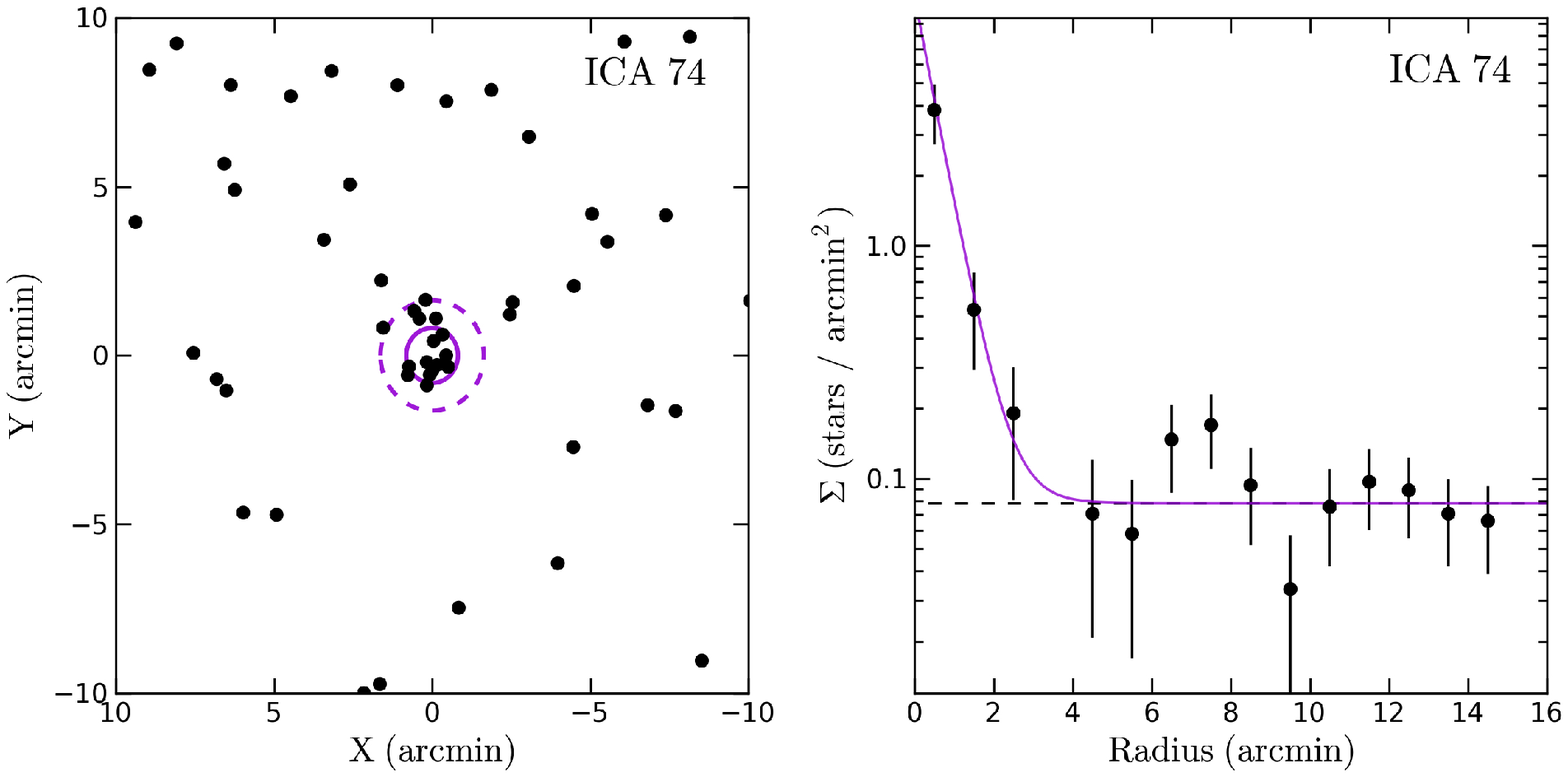} \\
\vspace{1mm}
\includegraphics[width=80mm]{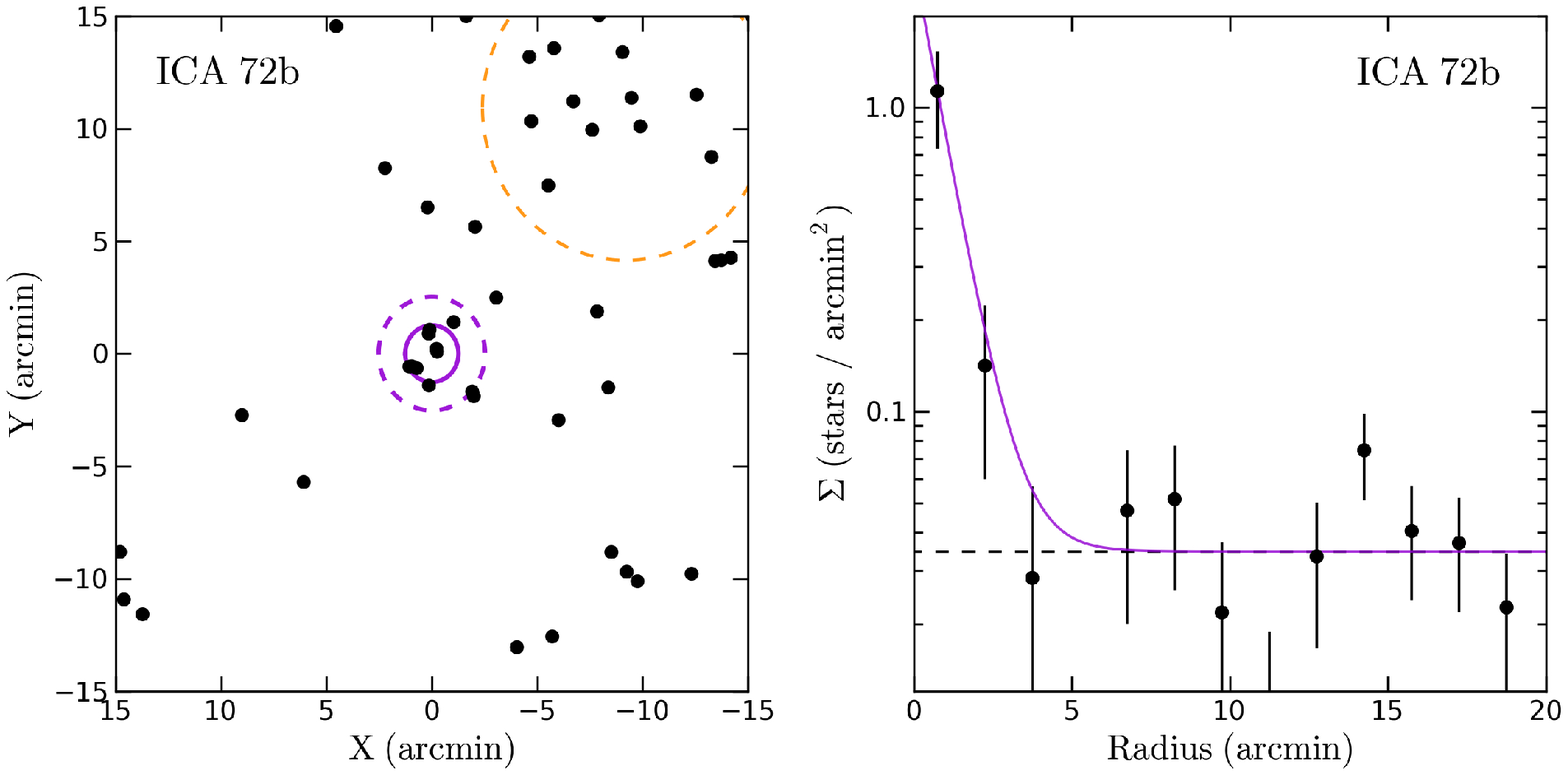}
\hspace{5mm}
\includegraphics[width=80mm]{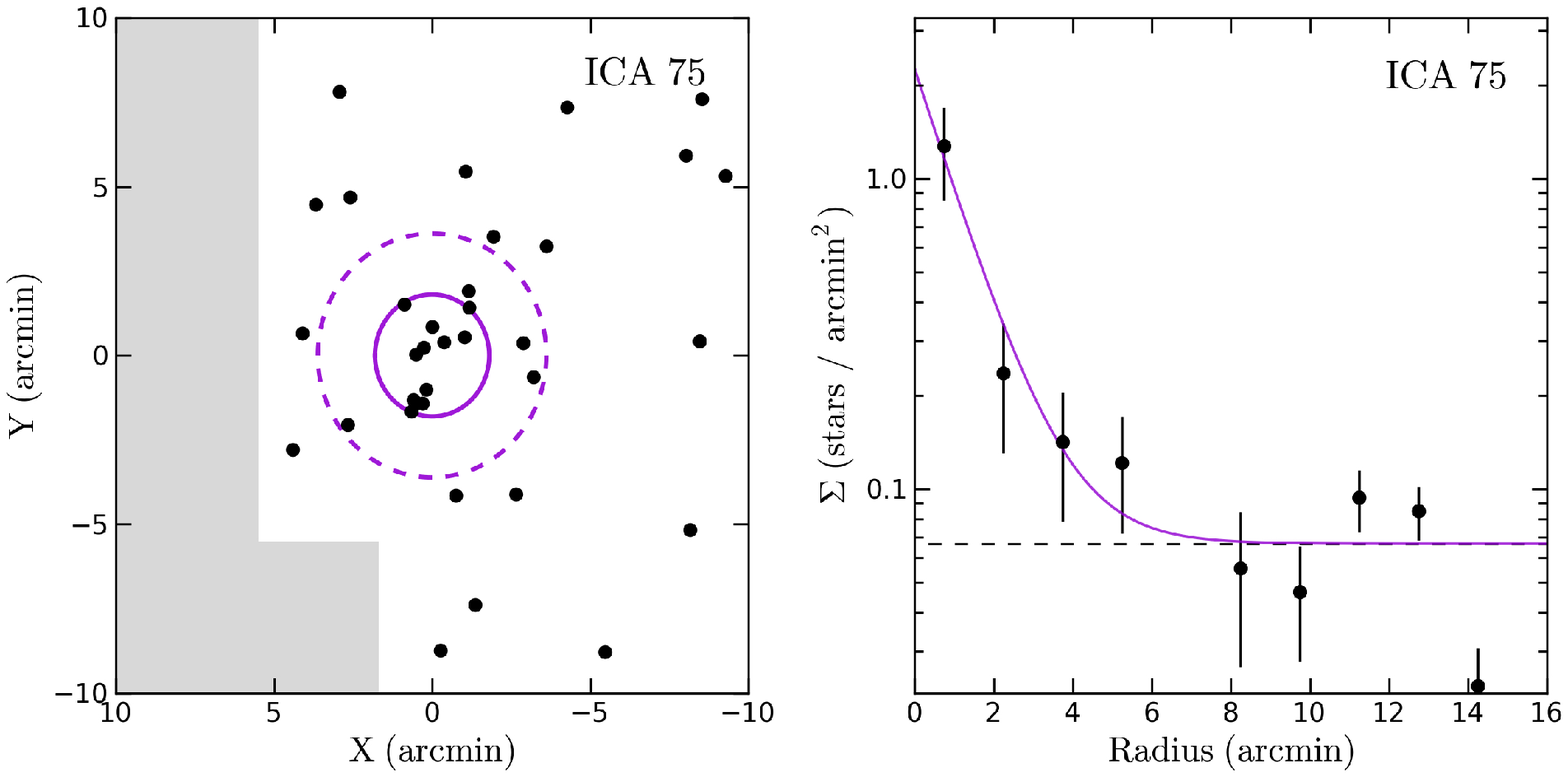}
\end{center}
\caption{Structural properties of the main young stellar associations in the eastern inter-Cloud region, measured using the maximum likelihood procedure described in the text \citep[see also][]{martin:08,martin:16}. Two panels are presented per target. The left panel shows the spatial distribution of stars falling in the CMD selection box for the region $\mathcal{A}$ surrounding the association. The projections are gnomonic, with the origin corrected for the derived offsets $(x_0,y_0)$. Also marked are ellipses with semi-major axis $r_h$ (solid line) and $2r_h$ (dashed line), ellipticity $e$, and orientation $\theta$ east of north. Shaded areas indicate regions that were not imaged in our survey, and which were excluded from the calculation of the background level $\Sigma_b$.  The right panel for each target shows the best-fitting exponential density profile as a function of elliptical radius $r$. The data are marked with solid points (each of which has Poissonian error bars), and the background level is indicated with the dashed horizontal line. We could not obtain a solution for the putative association ICA 72; the location of this structure is circled with a yellow dashed line in the left-hand panel for ICA 72b.
\label{f:youngstruct}}
\end{figure*}

Table \ref{t:struct} presents the measured structural parameters for each association; the results are also shown graphically in Figure \ref{f:youngstruct}. The derived positional offsets $(x_0,y_0)$ are not provided explicitly, but are incorporated into the listed coordinates for each target in Table \ref{t:params}, where we have taken the base positions from \citet{battinelli:92}, converted these to J2000.0, and then shifted them by $(x_0,y_0)$. As noted above, we could not obtain a solution for the object catalogued by \citet{battinelli:92} as ICA 72 -- the reason for this can be seen in the relevant panel of Figure \ref{f:youngstruct} where the putative structure is circled in yellow.  If there is an association at this location then it is very diffuse with a central surface density barely above the background level.

\begin{figure*}
\begin{center}
\includegraphics[height=83mm]{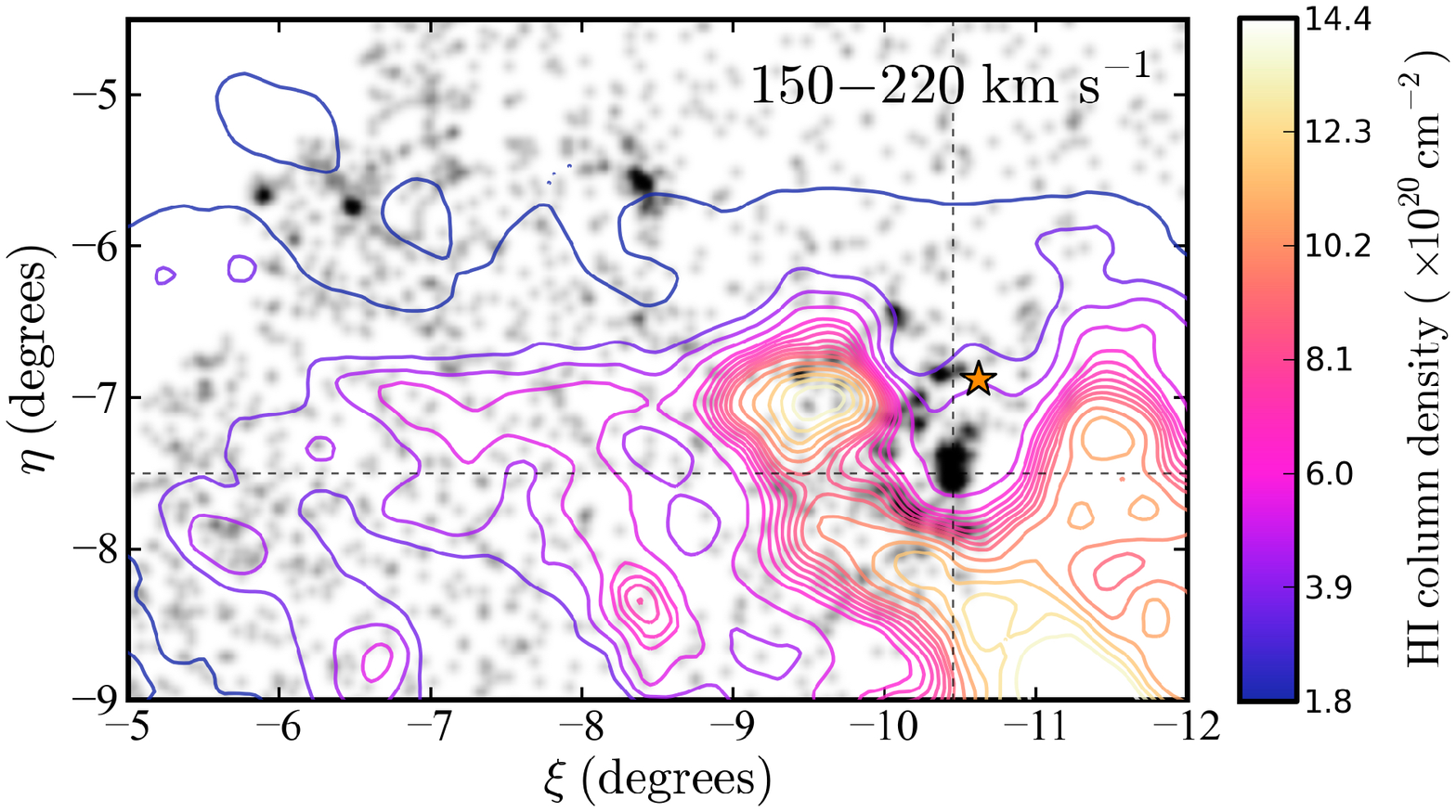}
\end{center}
\caption{Spatial density map of young stars in the eastern half of the inter-Cloud region, together with H{\sc i} column density contours from the Parkes Galactic All-Sky Survey \citep[GASS;][]{naomi:09,kalberla:10,kalberla:15} for radial velocities in the range $v_{\rm LSR} = 150-220$\ km$\,$s$^{-1}$. A low column density hole in the H{\sc i} distribution is evident, aligned perfectly with the stellar core-shell structure. The vertical and horizontal dashed lines indicate the locations of the perpendicular slices through the GASS data cube plotted in Figure \ref{f:slices}.  The position of the young star DI 1388 is marked with a yellow point in the upper portion of the hole (see the discussion in Section \ref{s:discussion}).
\label{f:gas}}
\end{figure*}
  
The two most populous targets -- ICA 73 and the core at the centre of the shell-like structure -- have remarkably large half-light radii of order $\sim 100$\ pc. The remaining four associations (ICA 72a, 72b, 74 and 75), for which we obtained constrained solutions with the ellipticity set to zero, also have sizeable half-light radii -- even the most compact of these (ICA 74, with $r_h = 12.5$\ pc) is many times larger than is typically seen for young clusters and associations in the Magellanic Clouds \citep[$\approx 1-2$\ pc,][]{mackey:03a,mackey:03b,mackey:08}. A similar observation was made by \citet{bica:15} for young associations in the western half of the inter-Cloud  region. Returning to ICA 73 and the core cluster, it is also notable that these two systems are both rather elliptical with $e\approx0.5$, and have position angles that match to within $\sim 15\degr$. Intriguingly these are also both closely aligned, to better than $\sim 10\degr$, with the position angle of the LMC relative to the SMC (which is $\theta\approx48\degr$). 

To estimate an integrated luminosity and total stellar mass for each association, we follow a procedure similar to that outlined by \citet{martin:16}. For a given association we assume a PARSEC isochrone of the appropriate age and shifted to the measured distance modulus, together with a \citet{kroupa:01} initial mass function (IMF) over the mass range defined by the isochrone. Next, we randomly draw a target number of stars $N_*$ from the structural parameter MCMC chain for the association. To build a model cluster we randomly sample the IMF using the isochrone to determine the $g$- and $r$-band flux per star, and add these fluxes to the running integrated luminosity and total mass. After each star is generated, we test whether it falls into the CMD selection box defined in Figure \ref{f:cmdmap}. Once we have flagged $N_*$ stars as falling in the selection box, the model is complete. We use the photometric transformation defined by Lupton on the SDSS web pages\footnote{\href{http://www.sdss.org/dr13/algorithms/sdssUBVRITransform/\#Lupton2005}{www.sdss.org/dr13/algorithms/sdssUBVRITransform/\#Lupton2005}} to convert the integrated $g$- and $r$-band magnitudes to Johnson $V$. 

We generate $10^5$ model realisations per association. We define the integrated luminosity $M_V$ and the total mass $M_{*}$ for a given association as the $50^{\rm th}$ percentiles of the distributions of $10^5$ luminosities and masses respectively. The $1\sigma$ uncertainties are given by the $16^{\rm th}$ and $84^{\rm th}$ percentiles. We adopted this methodology because the distribution of luminosities, in particular, can be very asymmetric due to the presence of one or more very bright evolved stars in some models. This is rare for associations where $N_*$ is relatively small, but occurs in a third to a half of all models for the more luminous systems studied here. 

To estimate the luminosity and mass of the shell structure, we counted the number of stars in the CMD in Figure \ref{f:youngcmds} that fall in the usual selection box, and calculated the area of sky over which these stars were distributed. We then assumed the background surface density measured from the structural parameter fits for the cluster at the core of the shell, and subtracted this level from the overall star count to leave $N_* = 159 \pm 14$ for the shell (where the uncertainty is entirely due to that in the assumed background density, which in turn comes from the uncertainty in $N_*$ for the core). With this estimate in hand, it was straightforward to generate random models as described above to determine $M_V$ and $M_{*}$. 

The results of our measurements are presented in Table \ref{t:struct}. In general the young stellar associations in our survey footprint are very low luminosity systems: ICA 72a, 72b, 74, and 75 have $M_V$ between $-2$ and $-3$ and $M_{*} \sim 100-200\,{\rm M}_\odot$. ICA 73 and the core and shell structures are considerably more substantial, each with $M_V$ between $-4.5$ and $-6$ and $M_{*} \sim 500-1200\,{\rm M}_\odot$ -- although, as previously noted, the shell appears fragmented into at least a dozen significant concentrations of young stars. On average, each of these fragments has a luminosity much more akin to the smaller isolated systems we have studied (ICA 72a, 72b, 74, and 75). Speculatively, this may reflect the characteristic scale of star formation in this part of the inter-Cloud region; if so, it is also possible that the more massive core and ICA 73 associations were formed via the merger of several such fragments (which could also help explain their much larger half-light radii). Nonetheless, it is worth emphasising that, when considered as a single structure, the shell has the highest luminosity of all the associations studied here -- even greater than the cluster located at its centre.

\subsection{Relationship to the Magellanic Bridge of H{\sc i}}
\label{ss:gas}
It is informative to compare the locations of the main concentrations of young stars in the inter-Cloud region to the distribution of H{\sc i} gas. It is well known that the young populations between the LMC and the SMC trace the Magellanic H{\sc i} Bridge quite closely \citep[e.g.,][]{dinescu:12,skowron:14,belokurov:17}; however our new deep photometry facilitates a much more detailed comparison than has previously been possible.

The H{\sc i} gas in the Magellanic Bridge is well separated from the Galactic foreground in velocity space, with a total span of $v_{\rm LSR} \approx 150-300$\ km$\,$s$^{-1}$ in the local-standard-of-rest frame \citep[see e.g.,][]{bruns:05,nidever:08,nidever:10}. By examining different velocity intervals across this range, we discovered clear signatures of interplay between the young stars and gas. In Figure \ref{f:gas} we reproduce our density map of young stars in the eastern inter-Cloud region, but now overplot contours of H{\sc i} column density from the Parkes Galactic All-Sky Survey \citep[GASS;][]{naomi:09,kalberla:10,kalberla:15} for velocities in the interval $v_{\rm LSR} = 150-220$\ km$\,$s$^{-1}$. While the young associations ICA 73, 74, and 75 show no obvious correspondence with the H{\sc i} at these velocities, there is a striking correlation between the gas and young stars in the region occupied by the core-shell feature. Specifically, there is a relatively low column density ``hole'' in the H{\sc i} distribution centred approximately on the cluster at the core of the stellar structure. The hole is bounded on all sides by some of the densest H{\sc i} features in the inter-Cloud region, except in the direction of increasing $\eta$ where the column density slowly declines. The shell of young stars matches precisely the location of the transition between low and high column densities at the edges of the H{\sc i} hole. 

\begin{figure}
\begin{center}
\includegraphics[width=0.48\textwidth]{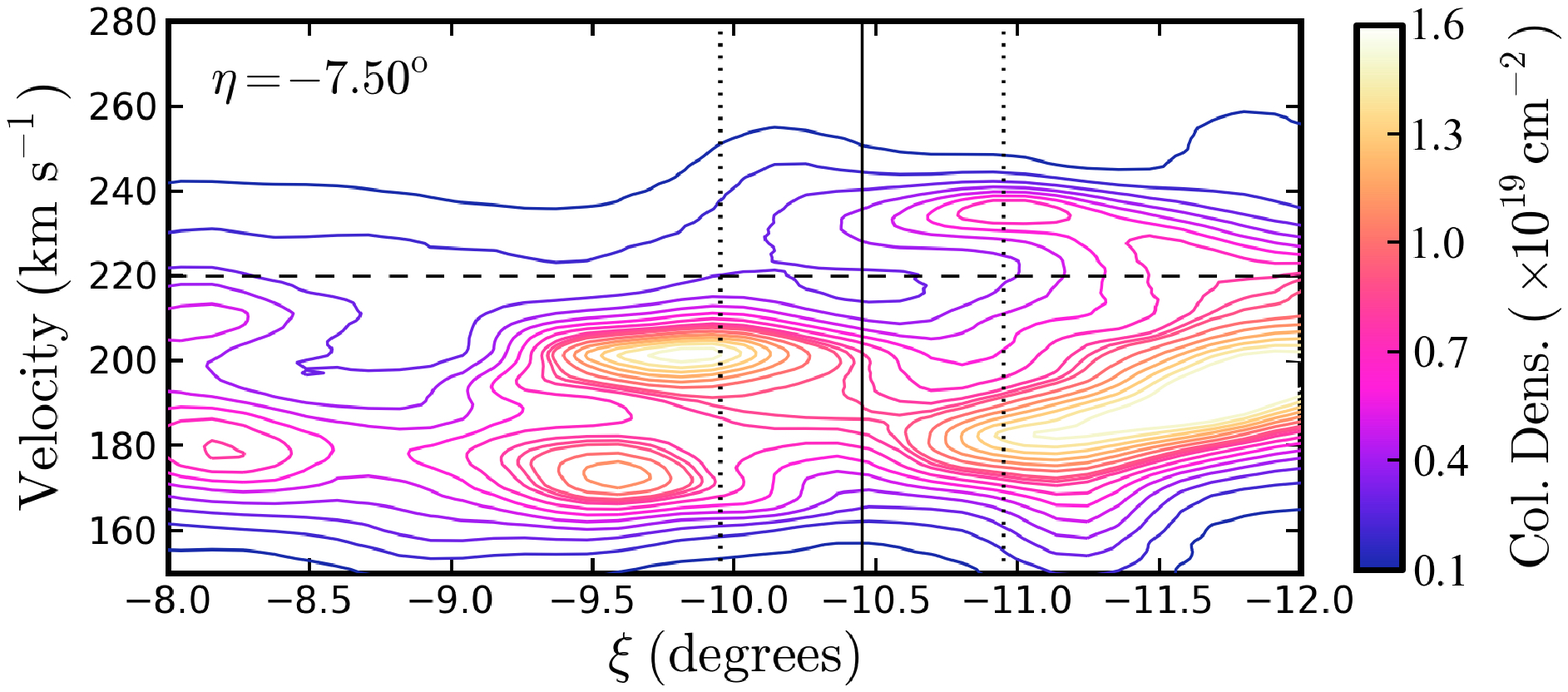}\\
\includegraphics[width=0.48\textwidth]{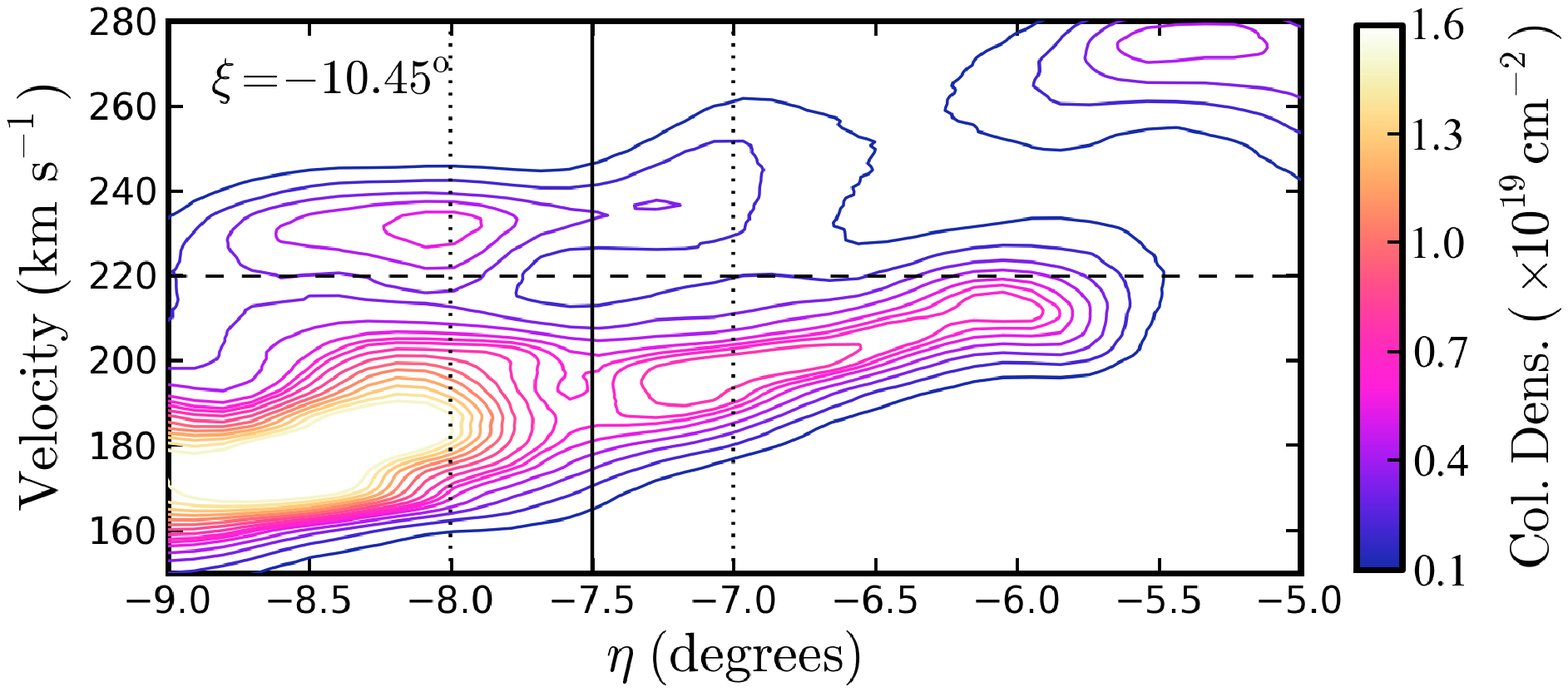}\\
\end{center}
\caption{H{\sc i} position-velocity profiles showing perpendicular slices through the GASS data cube at the location of the core association (i.e., along the lines of constant $\xi = -10.45$ and $\eta = -7.50$ as marked in Figure \ref{f:gas}).  The vertical black lines show the position of the core cluster -- the solid line marks its centre and the dotted lines at $\pm 0.5$ degrees indicate the approximate angular extent of the evacuated bubble. The black horizontal dashed line sits at $v_{\rm LSR} = 220$\ km$\,$s$^{-1}$; gas with velocity in the range $v_{\rm LSR} = 150-220$\ km$\,$s$^{-1}$ exhibits a strong correlation with the young stellar populations (again as shown in Figure \ref{f:gas}), while that with velocity in the range $v_{\rm LSR} = 220-280$\ km$\,$s$^{-1}$ shows little correlation (as in Figure \ref{f:gas2}). At the location of the bubble the H{\sc i} velocity profile has peaks near $v_{\rm LSR} \approx 180$\ km$\,$s$^{-1}$ and $\approx 205$\ km$\,$s$^{-1}$ -- i.e., a typical width of $\sim 25$\ \ km$\,$s$^{-1}$.
\label{f:slices}}
\end{figure}
 
Based on this map, it is reasonable to hypothesize that stellar winds and supernovae from massive stars in the central cluster swept up the surrounding H{\sc i} gas to produce an expanding evacuated ``bubble'' with material piled up on its surface, triggering a shell of new star formation. Other examples of sequential star formation have been observed in the LMC \citep[e.g.,][]{oey:95,oey:98}. It is possible that the bubble has ``blown out'' via the lower column density region in the direction of increasing $\eta$, allowing the hot interior gas to vent.

We investigate this scenario in more detail in Section \ref{ss:feedback}; for now, we show in Figure \ref{f:slices}, perpendicular slices through the GASS data cube at the location of the core association (i.e., along the lines of constant $\xi = -10.45$ and $\eta = -7.50$ as marked in Figure \ref{f:gas}). Considering only the H{\sc i} with $v_{\rm LSR} = 150-220$\ km$\,$s$^{-1}$, the velocity profile exhibits a double-peaked structure; this is especially evident in the top panel of Figure \ref{f:slices}. The peaks sit near $v_{\rm LSR} \approx 180$\ km$\,$s$^{-1}$ and $\approx 205$\ km$\,$s$^{-1}$ -- i.e., the typical width across the region occupied by the evacuated bubble is $\sim 25$\ km$\,$s$^{-1}$ across the region occupied by the evacuated bubble. Under the assumption that this can be attributed entirely to expansion of the shell, it indicates an expansion velocity of $v_{\rm exp} \approx 12.5$\ km$\,$s$^{-1}$.

\begin{figure*}
\begin{center}
\includegraphics[height=83mm]{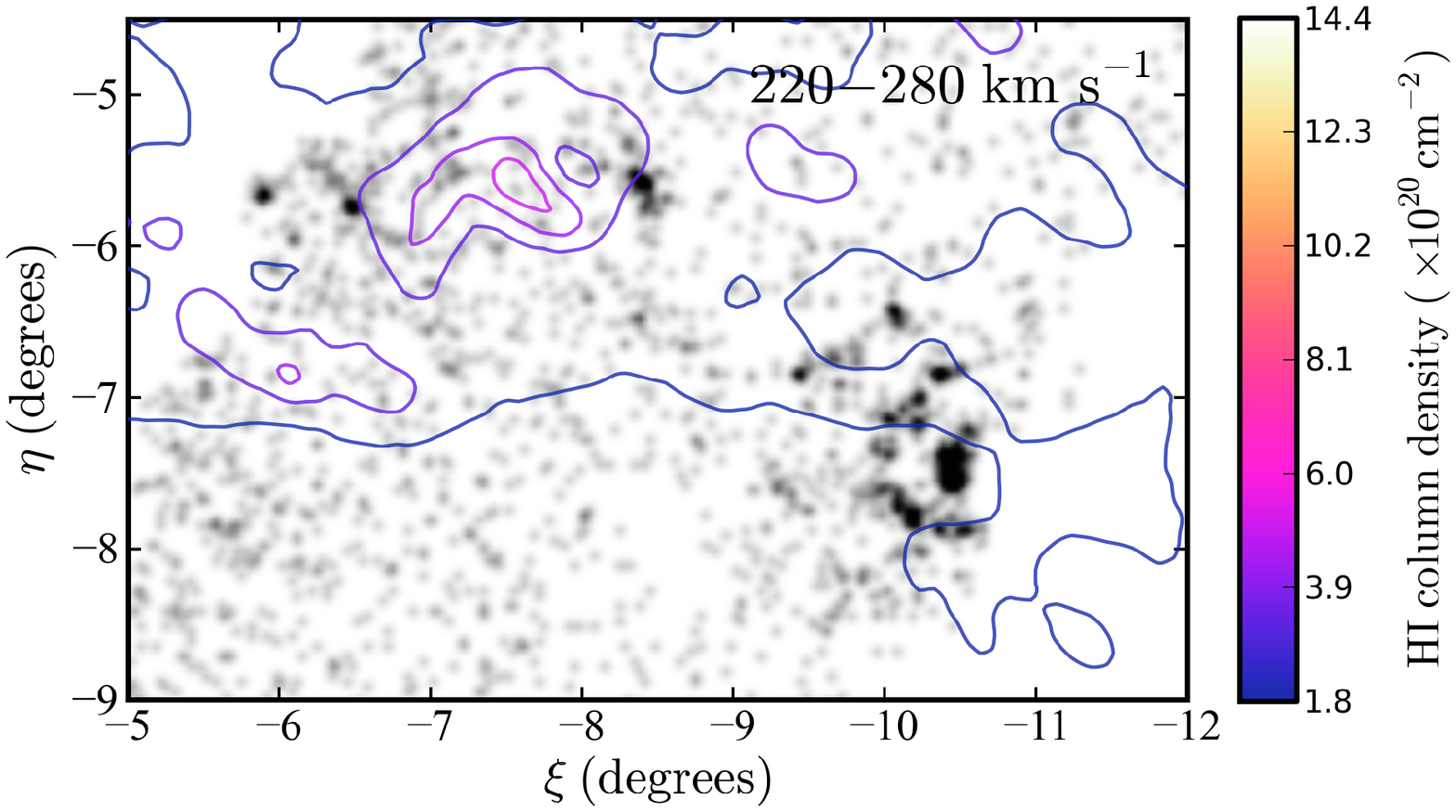}
\end{center}
\caption{Spatial density map of young stars in the eastern half of the inter-Cloud region, together with H{\sc i} column density contours from GASS for radial velocities in the range $v_{\rm LSR} = 220-280$\ km$\,$s$^{-1}$. This higher velocity H{\sc i} is more tenuous and possesses a more northerly distribution than the gas shown in Figure \ref{f:gas}. In addition, there is a possible correlation between the highest density H{\sc i} clump and the location of the young associations ICA 73 and 74.
\label{f:gas2}}
\end{figure*}

The position-velocity slices in Figure \ref{f:slices} also show evidence for a gaseous component with velocity in the range $v_{\rm LSR} \sim 220-280$\ km$\,$s$^{-1}$. To explore the origin of this we show, in Figure \ref{f:gas2}, column density contours for H{\sc i} in this velocity interval overplotted on our density map of young stars. At these velocities the distribution of the gas is quite different than for the slice plotted in Figure \ref{f:gas}.  Whereas the lower velocity H{\sc i} has higher column densities and is concentrated to the south of our survey region, the higher velocity gas is much more tenuous and preferentially located in the north. This matches previous observations by \citet{muller:04} who suggested that the Magellanic Bridge may be a projection of two kinematically and morphologically distinct gaseous features. The H{\sc i} distribution shown in Figure \ref{f:gas2} shows little apparent correlation with the main young stellar associations, apart from the highest density peak which sits neatly between ICA 73 and 74. Radial velocity measurements for these two associations would be needed to test whether or not this is a chance projection.

\section{Discussion}
\label{s:discussion}
Our new deep imaging of the eastern half of the Magellanic inter-Cloud region has allowed us to study the properties of the young stellar populations in this location with unprecedented clarity. Because these stars have recently formed out of gas in the Magellanic Bridge, they are valuable independent probes for exploring the characteristics and origin of this structure. They are also of considerable intrinsic interest, due to the low density and low metallicity nature of the star formation, and its location in the outer Galactic halo where the tidal force of the Milky Way is relatively benign.

\subsection{Distribution of young stellar populations}
As have many previous studies of this region, we find that the young stars are predominantly concentrated into a number of associations distributed between the LMC and SMC. Our colour-magnitude diagrams for the main young associations in the eastern inter-Cloud region indicate that these stellar systems are generally younger than $\approx 30$\ Myr (although the easternmost objects ICA 74 and 75 could potentially be as old as $\sim 200$\ Myr), and sit at distances intermediate between those of the LMC and SMC. This is consistent with the favoured scenario where the Magellanic Bridge of H{\sc i} gas has been largely stripped from the SMC due to a close interaction $\sim 200$\ Myr ago \citep[e.g.,][]{gardiner:96,besla:12}. Under the assumption that the young associations all share similar metallicities comparable to that observed for the gas in the Bridge -- i.e., $[$M$/$H$] \approx -1$ -- we find that they sit, on average, $\sim 4$\ kpc further along the line-of-sight than the LMC centre. Our measurements allow for a scenario where all the associations have the same line-of-sight distance, or for one where there is a small line-of-sight depth (of $\approx 1.25$\ kpc) to the overall distribution of young populations. We can exclude the possibility of any much greater variations in the distances to individual associations. Previous studies have typically obtained similar results for young inter-Cloud associations further to the west (i.e., closer to the SMC) than those studied here \citep[e.g.,][]{grondin:92,demers:98}; however, see \citet{bica:15} for an alternative view. 

If there {\it is} a non-zero line-of-sight depth to the distribution of young stellar populations in our survey region, it could be the result of a simple distance gradient along the Bridge, with clusters projected closer to the SMC (in our case, the core-shell structure) lying at greater distances than those projected closer to the LMC (in our case the ICA 72 complex, ICA 73, 74 and 75). On the other hand, work by \citet{muller:04} revealed an apparent discontinuity in the velocity of the H{\sc i} gas in the Bridge, with the higher velocity and less turbulent gas sitting to the north of the axis joining the LMC and SMC, and the lower velocity and more turbulent gas lying to the south. These authors concluded that the Magellanic Bridge might be a projection of two kinematically and morphologically distinct filaments, possibly representing two distinct arms of gas emanating from the SMC. Absorption-line studies of the gas in the Magellanic Bridge commonly find different components with different velocities, different abundance patterns, and even different ionization fractions along the same line-of-sight \citep[e.g.,][]{lehner:02,lehner:08,misawa:09}.

From Figures \ref{f:gas} and \ref{f:gas2} we have observed that the lower velocity H{\sc i} gas in the portion of the Bridge that we have imaged, with $v_{\rm LSR} \la 220$\ km$\,$s$^{-1}$, is clearly associated with the core-shell structure seen at the western edge of our survey footprint but almost absent from the region around ICA 73, 74, and 75 to the east. On the other hand, the more tenuous higher velocity gas, with $v_{\rm LSR} \ga 220$\ km$\,$s$^{-1}$, shows no apparent correlation with the core-shell structure, but has its highest density peak projected precisely between ICA 73 and 74.  The fact that these stellar associations are much smaller than the scale of the H{\sc i} clump may suggest that this is a chance alignment. If, however, the stars and gas are indeed correlated in this region, it could plausibly indicate that the associations at the eastern end of the Bridge were formed from gas in the higher velocity filament. Radial velocity measurements for stars in each of the young associations we have studied would be valuable for testing this scenario. In addition, abundance measurements \citep[as in, e.g.,][]{dufton:05,hunter:07,trundle:07} would (i) allow us to test whether associations in the putative higher velocity gas filament have systematically different abundance patterns to those in the lower velocity filament; (ii) allow us to check for cluster-to-cluster abundance variations which could provide information on how well mixed the gas in the Magellanic Bridge is; and (iii) help fix the absolute distances to the young associations\footnote{In Section \ref{ss:properties} we found that isochrones up to $\sim 0.5$ dex more metal-rich or metal-poor than $[$M$/$H$] = -1$ fit the {\it shape} of the main sequence equally well but shift the measured distance moduli by up to $\pm 0.2$ mag.}.

\subsection{Stellar feedback in the Magellanic Bridge}
\label{ss:feedback}
The most intriguing young stellar structure in the eastern inter-Cloud region is the core-shell feature revealed at high contrast by our deep observations. Based on the morphology and velocity profile of the coincident H{\sc i}, which shows evidence for an expanding low density evacuated region with edges that are perfectly aligned with the arc of young stars, we hypothesized that stellar winds and supernovae from massive stars in the central cluster swept up and compressed the surrounding gas into a shell, triggering a new burst of star formation. Similar examples of H{\sc i} holes and shells are commonly seen in disk environments, including the Milky Way, other Local Group galaxies, and more distant systems \citep[see e.g.,][and references therein]{naomi:02,boomsma:08}. The present case is unusual, however, in that the central power source is still visible and its properties can be measured quite precisely. This allows us to use different methods from the literature to estimate the energy required to drive the formation of the evacuated bubble, and then check their consistency by comparing the results to the available budget given the inferred number of massive stars present in the central cluster.

Perhaps the simplest way to estimate the kinetic energy $E_k$ necessary to form the bubble is by calculating the mass $M_s$ of H{\sc i} swept into the shell:
\begin{equation}
E_k = 0.5 \, M_s \, v_{\rm exp}^{2} = 4\times10^{43} \pi \, R^2 \, \Sigma \, v_{\rm exp}^{2}\,\,\,\,{\rm erg.}
\label{e:kinetic}
\end{equation}
Here $R$ is the observed radius of the shell in pc, $v_{\rm exp}$ is the observed expansion velocity in km$\,$s$^{-1}$, and $\Sigma$ is the surface mass density of the ambient gas into which the bubble expanded.  We have already measured $R \approx 480$\ pc, and $v_{\rm exp} \approx 12.5$\ km$\,$s$^{-1}$, while the column density at the position of the bubble is $\approx 5\times 10^{20}$\ cm$^{-2}$, which corresponds to a surface mass density of $4\,{\rm M}_\odot\,{\rm pc}^{-2}$ (see Figure \ref{f:gas}, although this should probably be considered a lower limit given the substantially higher density in most of the immediately adjacent locations). The resulting shell mass is $M_s \approx 1.1\times 10^7\,{\rm M}_\odot$, while the inferred kinetic energy is $E_k \approx 1.8\times 10^{52}$\ erg. The uncertainty in our estimate for $v_{\rm exp}$ is dominant in this calculation; adopting an uncertainty of $20\%$ (i.e., assuming $v_{\rm exp}$ falls in the range $\approx 10-15$\ km$\,$s$^{-1}$ as per Figure \ref{f:slices}) leads to an uncertainty of $\pm 0.7\times 10^{52}$\ erg in the kinetic energy.

It is also possible to define an energy $E_E$, which is the equivalent energy that would have to be instantaneously deposited at the centre of the shell to account for the observed radius and expansion velocity \citep[see][]{heiles:84}. The literature contains a variety of methods for estimating this quantity, although we note that the approximation that the energy is instantaneously deposited is likely not appropriate for our situation (where we are considering an evolving OB association rather than a single supernova explosion). Following \citet{chevalier:74}, who conducted hydrodynamical simulations that included radiative cooling, we first try:
\begin{equation}
E_E = 5.3\times10^{43}\, n_0^{1.12}\, R^{3.12}\, v_{\rm exp}^{1.40} \,\,\,\,{\rm erg}
\label{e:chevalier}
\end{equation}
\citep[see also, e.g.,][]{heiles:79,naomi:02}. In this expression, $n_0$ is the initial ambient density in cm$^{-3}$ of the gas. To estimate this we start with our adopted column density of $\approx 5\times 10^{20}$\ cm$^{-2}$, and assume that the gas is at least as deep along the line-of-sight as the present radius of the shell (otherwise the hot gas would have vented in this direction and stalled the expansion at a smaller size). This leads to $n_0 \approx 0.17$\ cm$^{-3}$, and an expansion energy of $E\approx 5.8\times 10^{52}$ erg. Adopting uncertainties of $10\%$, $5\%$, and $20\%$ in our measurements of $n_0$, $R$, and $v_{\rm exp}$ respectively, yields an overall uncertainty of $\pm 2.0\times 10^{52}$ erg in this calculation.

An alternative estimate of the expansion energy comes from \citet*{cioffi:88}:
\begin{equation}
E_E = 6.8\times10^{43}\, n_0^{1.16} R^{3.16}\, v_{\rm exp}^{1.35}\, \zeta_m^{0.16} \,\,\,\,{\rm erg.}
\label{e:cioffi}
\end{equation}
Again, this expression is derived from hydrodynamical simulations that included radiative cooling, and again it makes the approximation that the energy is deposited instantaneously. The formula has a similar functional form to that of Eq. \ref{e:chevalier} but incorporates an explicit metallicity dependence, $\zeta_m$, which is the abundance of the gas relative to solar. Taking $[$M$/$H$] = -1.0$ for the Magellanic Bridge, such that $\zeta_m \approx 0.1$, we find that for the observed shell radius of $480$\ pc, expansion velocity $v_{\rm exp} \approx 12.5$\ km$\,$s$^{-1}$, and the assumed ambient gas density $n_0 \approx 0.17$\ cm$^{-3}$, the expansion energy $E_E\approx 5.4\times10^{52}$\ erg. Adopting uncertainties of $10\%$, $5\%$, and $20\%$ in our measurements of $n_0$, $R$, and $v_{\rm exp}$ as before, and adding an uncertainty of $\sim 20\%$ in $\zeta_m$, yields an overall uncertainty of $\pm 1.8\times10^{52}$ erg. 

We have obtained three estimates for the energy required to drive the expansion of the observed bubble. The expressions of \citet{chevalier:74} and \citet{cioffi:88} take into account radiative energy losses, and thus in these cases the calculated energy of $\approx 5.5 \times 10^{52}$\ erg should be the total required input. On the other hand the first simple estimate of $\sim 1.8 \times 10^{52}$\ erg instead gives only the kinetic energy, which could be as little as $\sim 10\%$ of the total amount needed as most of the supernova explosion energy will ultimately be radiated \citep[e.g.,][]{thornton:98}. Under the assumption that a single supernova injects $\sim 10^{51}$\ erg into the surrounding ISM, our estimates suggest the total energy requirement is roughly equivalent to the expected input from $\ga 50$ OB stars.

Our simple random realisations of the star cluster at the core of the shell, from which we determined an integrated luminosity $M_V = -5.3^{+0.5}_{-1.1}$, indicate a present-day mass of $900^{+110}_{-100}\,{\rm M}_\odot$ in a system containing $1850 \pm 250$ stars \citep[recall that this assumes the IMF of][]{kroupa:01}. With an age of $\approx 30$\ Myr, the present-day main sequence turn-off mass is $\sim 8\,{\rm M}_\odot$. For a \citet{kroupa:01} IMF, the fraction of stars in a zero-age cluster with masses greater than $8\,{\rm M}_\odot$ is about $0.7\%$. This means that in the central cluster of the core-shell system, we expect that only $\sim 10-15$ stars have evolved off the main sequence and exploded as supernovae since its formation. This is discrepant by a factor of roughly $5$ with the number implied by our energy estimates.

This discrepancy could arise because the simple expressions for the expansion energy considered above are not strictly applicable to the system we are investigating. To explore this we consider the theory of superbubble expansion \citep[e.g.,][]{weaver:77,maclow:88}, which may be more appropriate here because it describes a supershell produced around an OB association. In particular, this model accounts for the collective effects of multiple sequential supernovae rather than making the assumption that all the energy is deposited instantaneously. As long as the superbubble has not blown out of the surrounding H{\sc i}, the predicted radius is \citep[e.g.,][]{mccray:87}:
\begin{equation}
R = 97 \, (N_{\rm SN} E_{51})^{0.2} \, n_0^{-0.2} \, t_7^{0.6}\,\,\,\,{\rm pc,}
\label{e:bubble1}
\end{equation}
while the expected velocity of the shell is given by:
\begin{equation}
v_{\rm exp} = 5.7 \, (N_{\rm SN} E_{51})^{0.2} \, n_0^{-0.2} \, t_7^{-0.4}\,\,\,\,{\rm km}\,{\rm s}^{-1}.
\label{e:bubble2}
\end{equation}
In these expressions $N_{\rm SN}$ is the number of stars that will become supernovae over the lifetime of the OB association, $E_{51}$ is the assumed energy of an individual supernova in units of $10^{51}$\ erg, and $t_7$ is the age of the association in units of $10$\ Myr. We assume $E_{51} \sim 1.0$ and that the age of the core cluster is $\approx 30$\ Myr. The number of stars that will become supernovae over the lifetime of the OB association is more difficult to determine. We have shown that $\sim 10-15$ stars have likely already exploded as supernovae in our cluster; how many more will explode in the future is dependent on what we assume the minimum mass of a supernova progenitor to be. Given that the present-day main sequence turn-off mass is $\sim 8\,{\rm M}_\odot$, which is close to the expected minimum mass for a Type II supernova, adopting $N_{\rm SN} = 15 \pm 5$ seems reasonable. In this case, we find the predicted radius and velocity of the supershell to be $R = 460 \pm 60$\ pc and $v_{\rm exp} \approx 9.0 \pm 1.0$\ km$\,$s$^{-1}$, respectively.

These estimates are close to the observed values for our shell. Its measured radius of $R\approx 480$\ pc sits well within the uncertainty on the theoretical prediction, while the expansion velocity of $v_{\rm exp} \approx 12.5$\ km$\,$s$^{-1}$ inferred from Figure \ref{f:slices} is only slightly larger than the predicted velocity. Importantly, the required number of supernova explosions is entirely consistent with the observed luminosity of the core association and a standard \citet{kroupa:01} IMF. 

An important caveat is that this model assumes the supernova rate is high enough to effectively provide a continuous injection of energy. If this requirement is not satisfied, the collective action of the supernovae can be substantially less efficient at driving the expansion of the bubble \citep[see e.g.,][and references therein]{vasiliev:17}. The fact that we observe good agreement between the predictions of the model and the observed properties of our bubble suggests that the central OB association is in the regime where the assumptions that Equations \ref{e:bubble1} and \ref{e:bubble2} are based on are valid; however this might require that the central association was more compact in the past (or at least that the massive stars were all centrally located).

In Section \ref{ss:gas} we noted the possibility that the bubble has vented to the north where the column density of the surrounding H{\sc i} is seen to be lowest. In this context it is interesting that the observed velocity of the shell is similar to the value predicted by Equation \ref{e:bubble2}, because it means that not much deceleration has taken place with respect to the prediction for a freely expanding superbubble. This would suggest that either the superbubble has not yet blown out or that if blow-out has occurred it was not long ago. It is also relevant that the observed expansion speed of $v_{\rm exp} \approx 12.5$\ km$\,$s$^{-1}$ is close to the sound speed in a gas at a temperature of $10^4\,$K ($c_s \sim 10$\ km$\,$s$^{-1}$), because the phase during which the bubble is sweeping out material stops once the expansion velocity drops to the speed of sound in the ambient gas. This could suggest that the expansion of the shell being studied here may nearly be over. One complication that is not accounted for by the theory is the presence of the subsequent generation of stars that has formed in the shell. We have shown that this association is $\sim 35\%$ more massive than that in the central cluster; it is therefore more than likely that this population is also contributing a significant amount of energy to the expansion, which may explain why the observed expansion velocity is slightly larger than that predicted by the theory. 

Finally, it is also worth considering results from previous, more direct studies of the gas properties towards the core-shell structure.  In particular, the young star DI 1388 is projected inside the evacuated bubble (see Figure \ref{f:gas}) and has been used as a target for absorption-line investigations. These have observed ``multiple gas phases in a complex arrangement'' along this line of sight in the Bridge \citep{lehner:02}.  About $70\%$ of the gas towards DI 1388 is ionized; moreoever two clear components are detected -- one at lower velocities ($165 \leq v_{\rm LSR} \leq 193$\ km$\,$s$^{-1}$) that is nearly fully ionized ($\sim 95\%$) and one at higher velocities ($193 \leq v_{\rm LSR} \leq 215$\ km$\,$s$^{-1}$) that is partially ionized ($\sim 50\%$) \citep{lehner:01,lehner:08}. Molecular hydrogen (H$_2$) is also seen along this sight line \citep{lehner:02}, but it is not clear how this may relate to the other components. Regardless of the complexities, these measurements directly indicate the presence of hot ionized gas coincident with the evacuated bubble spatially (at least in projection), and in radial velocity. It is also relevant that the maps of diffuse H$\alpha$ emission constructed by \citet{barger:13} exhibit a weak enhancement at this location (see their Figure 7, near $l,b \approx 291.8, -40.9$), providing independent evidence for warm ionized gas.

\subsection{Triggered star formation in the core-shell structure}
\begin{figure}
\begin{center}
\includegraphics[width=0.40\textwidth]{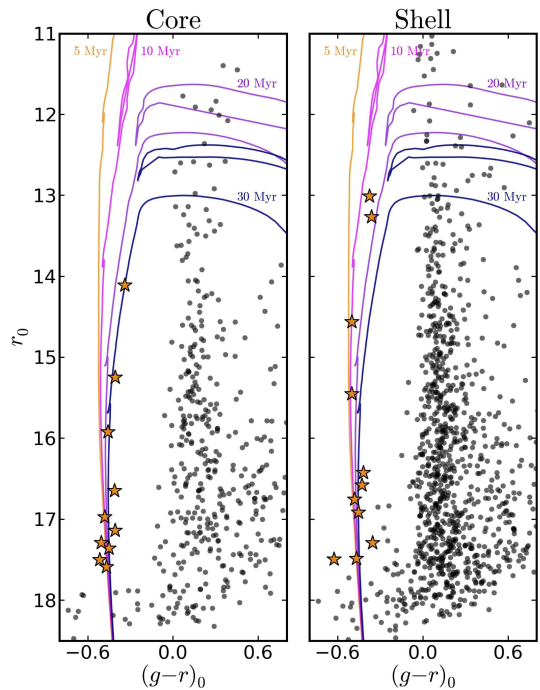}
\end{center}
\caption{Colour-magnitude diagrams for stars coincident with the core and shell structures, where the photometry has been taken from the SkyMapper Early Data Release \citep{wolf:16}. SkyMapper provides measurements for sources up to $\sim 4$ mag brighter than our {\it DECam} survey. Stars which have $r_0 < 17.7$ are considered to have ``reliable'' SkyMapper photometry in these fields, as indicated by the {\sc class\_star} parameter. Those that also have colour $(g-r)_0 < -0.25$ constitute the upper main sequence of the two stellar systems and are highlighted with yellow points. On each CMD we plot PARSEC isochrones with $[$M$/$H$] = -1$, distance modulus $\mu = 18.7$, and ages (upper to lower, as labelled) of $5$, $10$, $20$, and $30$\ Myr. These indicate that the shell structure is likely $\sim 10-15$\ Myr younger than the central cluster.
\label{f:skymapper}}
\end{figure}

If the scenario outlined above is correct, then stellar winds and supernova explosions due to massive stars in the central cluster are responsible for sweeping up the surrounding gas, compressing it, and triggering the formation of a new generation of stars in the shell. We might therefore expect the stars in the shell to be noticeably younger than those in the core. Given that our {\it DECam} observations saturate above $r \sim 15.5$, we were only able to place an upper age limit of $\sim 30$\ Myr on both the core and shell structures.  However, shallower photometry for this region has recently become available as part of the SkyMapper Early Data Release \citep[EDR,][]{wolf:16}\footnote{See also \href{http://skymapper.anu.edu.au}{skymapper.anu.edu.au}}; in Figure \ref{f:skymapper} we plot CMDs for stars coincident with the central core (left) and shell (right). These extend our range by at least four magnitudes to $r \approx 11.5$. As with our {\it DECam} observations, the SkyMapper data have been calibrated to the SDSS scale via APASS. Since the SkyMapper zero-points are uncertain at the few percent level due to various limitations in the EDR, we matched stars with those present in our {\it DECam} catalogues over the range $17.5 \la r_0 \la 16.0$ ($6$ stars for the core, and $5$ for the shell) to make small corrections to the zero-points. We plot PARSEC isochrones with $[$M$/$H$] = -1$, distance modulus $\mu = 18.7$, and ages of $5$, $10$, $20$, and $30$\ Myr on the CMDs. Although the core and shell structures each possess only a handful of very luminous stars, it is evident that those in the core are consistent with the $30$\ Myr isochrone, while those for the shell sit between the $10$ and $20$\ Myr tracks. This suggests that stars in the shell are likely to be $\sim 10-15$\ Myr younger than those in the core, consistent with the scenario where feedback from the central core has triggered star formation in the shell.

\subsection{Properties of the young stellar associations}
All of the young stellar associations that we have studied in the eastern inter-Cloud region appear to possess unusually extended structures with half-light radii ranging between $\sim 10-100$ pc. This contrasts strongly with young clusters in the LMC and SMC, which typically have half-light radii of $\sim 1-2$\ pc.  A similar pattern has been noted for associations in the western part of the inter-Cloud region by \citet{bica:15}. The low density of the young stellar associations in the inter-Cloud region might reflect the unusual conditions of star formation at this location. For example, \citet{elmegreen:08} has suggested that star formation in turbulent regions with relatively high Mach number and moderately low density can produce low density bound stellar systems (especially if the background tidal forces are low, to ensure survivability). Given that the gas in the Magellanic Bridge is thought to have been largely stripped from the SMC during a close encounter with the LMC $\approx 200$\ Myr ago, it is not difficult to imagine that some parts of the inter-Cloud region might satisfy these criteria. Indeed, we have noted that the position angles of the two most luminous young associations we have studied (ICA 73 and the cluster at the centre of the core-shell structure) are aligned with the axis joining the LMC and the SMC to better than $\approx 10\degr$. Since these systems are too young to have undergone dynamical relaxation, this must reflect the initial conditions and may suggest that star formation was triggered due to compression of the stripped Bridge gas during the close LMC-SMC encounter. 

It would be of considerable interest to use numerical $N$-body models to explore whether the young associations in the Magellanic Bridge will remain bound for a signficant period of time. This may be possible because they already appear to have largely expelled any residual gas, and moreover they now reside in the outer Milky Way halo where tidal forces are relatively benign. If we take our simple random realisations of the cluster at the centre of the core-shell association and fade the stellar populations to an age of $\sim 12$\ Gyr, the implied luminosity is $M_V \sim -1$. A similar calculation for ICA 73, which is presently not quite a factor of two less massive than the core cluster, yields a comparable result. These luminosities are extremely faint but match those for a number of ancient diffuse star clusters discovered in the outer Milky Way halo in recent years, some of which have half-light radii $r_h \ga 10$\ pc \citep[e.g.,][]{munoz:12,balbinot:13,kim:15}. Under the assumption that the diffuse structures seen for the young stellar associations in the inter-Cloud region are a signature of the particular star formation conditions prevalent in the Magellanic Bridge, and that this also holds for more massive clusters than those observed here, it is reasonable to speculate that the luminous extended globular clusters found in many dwarf galaxies and in the outskirts of larger galaxies \citep[see e.g.,][and references therein]{mackey:04,mackey:06,mackey:13,huxor:05,huxor:14,dacosta:09,hwang:11} could be tracers of gas-rich galaxy mergers and/or interactions that occurred in the early Universe.

\section{Summary}
We have used the {\it Dark Energy Camera} to conduct a deep contiguous imaging survey of the eastern half of the Magellanic inter-Cloud region. Our data allow us to explore the distribution and properties of the young stellar populations in this region with unparalleled clarity and resolution.  Our main results are:
\begin{enumerate}
\item{As observed by many previous studies, the young inter-Cloud populations stretch all the way from the SMC ``wing'' to the south-western outskirts of the LMC disk. They are strongly spatially clustered, forming a narrow chain of low-mass stellar associations. The young inferred ages of these associations ($\la 30$\ Myr for the largest in our survey), and the observation that they are projected on top of the densest regions of H{\sc i} seen in the Magellanic Bridge, strongly suggest that the young stars have formed {\it in situ} rather than having been stripped from the LMC or SMC. This conclusion is reinforced by the fact that we see clear evidence for stellar feedback altering the properties of the ambient gas in at least one location (see below).\vspace{1mm}}
\item{Under the assumption that the young associations share a similar metallicity to that typically measured for the gas in the Magellanic Bridge (i.e., $[$M$/$H$]\sim -1$), we find them to have distance moduli $\mu \approx 18.65-18.70$. This is intermediate between that for the LMC ($\mu = 18.49$) and SMC ($\mu = 18.96$). Our measurements do not allow for very large, random, cluster-to-cluster variations in $[$M$/$H$]$, nor do they allow for a line-of-sight depth or distance gradient greater than $\Delta\mu \approx 0.15$ across the eastern half of the inter-Cloud region. At low significance the data are consistent with a mild distance gradient such that the objects that are closer to the LMC in projection have shorter line-of-sight distances by $\approx 0.05$ mag; however we cannot rule out that all the associations sit at the same distance. It is reasonable to assume that the distances we measure for the young associations reflect that for the majority of the gas in this part of the Magellanic Bridge.\vspace{1mm}}
\item{The seven main associations in the surveyed area have masses in the range $\approx 100-1200\,{\rm M}_\odot$. They are also remarkably diffuse -- we find half-light radii as large as $\sim 100$\ pc for two of the most populous clusters. It is possible that the extended structures that we observe for the young stellar associations in the eastern inter-Cloud region may reflect the low-density conditions of star formation at this location. It is also notable that the two most populous clusters are both rather elliptical (with $e\approx 0.5$) and oriented such that their major axes are aligned with each other, and the position angle of the LMC relative to the SMC, to within $\sim 10-15\degr$. Since these systems are too young to have undergone dynamical relaxation, this must reflect the initial conditions and may suggest that star formation was triggered due to compression of gas stripped from one or other of the Clouds during their most recent close encounter.\vspace{1mm}}
\item{We identify a vast shell of young stars surrounding a compact core association, that is spatially coincident with a low column density ``bubble'' in the H{\sc i} distribution for velocities in the range $v_{\rm LSR} = 150-220$\ km$\,$s$^{-1}$. This structure has a radius of $R\approx 480$\ pc, and slices through the GASS data cube suggest the gaseous shell is expanding with a velocity of $v_{\rm exp} \approx 12.5$\ km$\,$s$^{-1}$. We argue that stellar winds and supernova explosions from massive stars in the central cluster swept up and compressed the ambient gas, triggering a new burst of star formation. The theory of superbubble expansion developed by, e.g., \citet{weaver:77}; \citet{mccray:87}; and \citet{maclow:88}, closely predicts the observed properties of the shell given the expected number of massive stars at the core inferred from its observed luminosity. We measure the mass of stars formed in the shell to be $\sim 35\%$ greater than the mass of the central cluster, while CMDs for bright stars at this location suggest that the shell could be $\sim 10-15$\ Myr younger than the association at its core. It is notable that young stars in the shell are not uniformly distributed but instead concentrate into at least a dozen fragments, each with a luminosity more akin to the smaller isolated associations seen nearby. This structure constitutes a superb example of positive stellar feedback operating in a relatively isolated environment.}
\end{enumerate}

\section*{Acknowledgments}
ADM is grateful for support from an Australian Research Council (ARC) Future Fellowship (FT160100206).  ADM and GDC also acknowledge support from the ARC through Discovery Projects DP1093431, DP120101237, and DP150103294. NMM-G acknowledges the support of the ARC through grant FT150100024. MF acknowledges the support of a Royal Society - Science Foundation Ireland University Research Fellowship. The research leading to these results has received funding from the European Research Council under the European Union's Seventh Framework Programme (FP/2007-2013)/ERC Grant Agreement no. 308024. We thank the International Telescopes Support Office (ITSO) at the Australian Astronomical Observatory (AAO) for providing travel funds to support our 2016A {\it DECam} observing run.

This project has used public archival data from the Dark Energy Survey (DES) together with data obtained with the {\it Dark Energy Camera} ({\it DECam}), which was constructed by the DES collaboration.
Funding for the DES Projects has been provided by 
the U.S. Department of Energy, 
the U.S. National Science Foundation, 
the Ministry of Science and Education of Spain, 
the Science and Technology Facilities Council of the United Kingdom, 
the Higher Education Funding Council for England, 
the National Center for Supercomputing Applications at the University of Illinois at Urbana-Champaign, 
the Kavli Institute of Cosmological Physics at the University of Chicago, 
the Center for Cosmology and Astro-Particle Physics at the Ohio State University, 
the Mitchell Institute for Fundamental Physics and Astronomy at Texas A\&M University, 
Financiadora de Estudos e Projetos, Funda{\c c}{\~a}o Carlos Chagas Filho de Amparo {\`a} Pesquisa do Estado do Rio de Janeiro, 
Conselho Nacional de Desenvolvimento Cient{\'i}fico e Tecnol{\'o}gico and the Minist{\'e}rio da Ci{\^e}ncia, Tecnologia e Inovac{\~a}o, 
the Deutsche Forschungsgemeinschaft, 
and the Collaborating Institutions in the Dark Energy Survey. 
The Collaborating Institutions are 
Argonne National Laboratory, 
the University of California at Santa Cruz, 
the University of Cambridge, 
Centro de Investigaciones En{\'e}rgeticas, Medioambientales y Tecnol{\'o}gicas-Madrid, 
the University of Chicago, 
University College London, 
the DES-Brazil Consortium, 
the University of Edinburgh, 
the Eidgen{\"o}ssische Technische Hoch\-schule (ETH) Z{\"u}rich, 
Fermi National Accelerator Laboratory, 
the University of Illinois at Urbana-Champaign, 
the Institut de Ci{\`e}ncies de l'Espai (IEEC/CSIC), 
the Institut de F{\'i}sica d'Altes Energies, 
Lawrence Berkeley National Laboratory, 
the Ludwig-Maximilians Universit{\"a}t M{\"u}nchen and the associated Excellence Cluster Universe, 
the University of Michigan, 
{the} National Optical Astronomy Observatory, 
the University of Nottingham, 
the Ohio State University, 
the University of Pennsylvania, 
the University of Portsmouth, 
SLAC National Accelerator Laboratory, 
Stanford University, 
the University of Sussex, 
and Texas A\&M University. 

This work has made use of the AAVSO Photometric All-Sky Survey (APASS), funded by the Robert Martin Ayers Sciences Fund.

This research has used data from the SkyMapper Early Data Release (EDR). The national facility capability for SkyMapper has been funded through ARC LIEF grant LE130100104 from the Australian Research Council, awarded to the University of Sydney, the Australian National University, Swinburne University of Technology, the University of Queensland, the University of Western Australia, the University of Melbourne, Curtin University of Technology, Monash University and the Australian Astronomical Observatory. SkyMapper is owned and operated by The Australian National University's Research School of Astronomy and Astrophysics. The survey data were processed and provided by the SkyMapper Team at ANU. The SkyMapper node of the All-Sky Virtual Observatory is hosted at the National Computational Infrastructure (NCI).











\bsp	
\label{lastpage}
\end{document}